\documentclass[aps,prl,amsmath,reprint,superscriptaddress]{revtex4-2}

\usepackage{graphicx}
\usepackage{dcolumn}
\usepackage{bm}
\usepackage[colorlinks=true, linkcolor=blue, citecolor=blue, urlcolor=blue,breaklinks=true]{hyperref}
\usepackage[mathlines]{lineno}
\usepackage{amsthm,amssymb}
\usepackage{physics}
\usepackage{mathrsfs}
\usepackage{mathtools}
\usepackage{xcolor}
\usepackage[caption=false]{subfig}
\usepackage{amsmath}

\rmfamily

\begin{document}

\title{Scalable multiparty steering based on a single pair of entangled qubits}
\author{Alex Pepper}
\email{alex0pepper@gmail.com}
\affiliation{Centre for Quantum Dynamics and Centre for Quantum Computation and Communication Technology (CQC$\,^2\!$T), Griffith University, Yuggera Country, Brisbane, 4111 Australia}
\author{ Travis J. Baker}
\email{dr.travis.j.baker@gmail.com}
\affiliation{Centre for Quantum Dynamics and Centre for Quantum Computation and Communication Technology (CQC$\,^2\!$T), Griffith University, Yuggera Country, Brisbane, 4111 Australia}
\affiliation{Nanyang Quantum Hub, School of Physical and Mathematical Sciences, Nanyang Technological
University, Singapore 637371}
\author{Yuanlong Wang}
\affiliation{Centre for Quantum Dynamics and Centre for Quantum Computation and Communication Technology (CQC$\,^2\!$T), Griffith University, Yuggera Country, Brisbane, 4111 Australia}
\affiliation{Key Laboratory of Systems and Control, Academy of Mathematics and Systems Science, Chinese Academy of Sciences, Beijing 100190, People’s Republic of China}
\author{Qiu-Cheng Song}
\affiliation{Centre for Quantum Dynamics and Centre for Quantum Computation and Communication Technology (CQC$\,^2\!$T), Griffith University, Yuggera Country, Brisbane, 4111 Australia}
\author{Lynden. K. Shalm}
\author{Varun B. Verma}
\author{Sae Woo Nam}
\affiliation{National Institute of Standards and Technology, 325 Broadway, Boulder, Colorado 80305, USA}
\author{Nora Tischler}
\email{n.tischler@griffith.edu.au}
\author{Sergei Slussarenko}
\author{Howard M. Wiseman}
\email{h.wiseman@griffith.edu.au}
\author{Geoff. J. Pryde}
\affiliation{Centre for Quantum Dynamics and Centre for Quantum Computation and Communication Technology (CQC$\,^2\!$T), Griffith University, Yuggera Country, Brisbane, 4111 Australia}

\date{\today}

\begin{abstract}

The distribution and verification of quantum nonlocality across a network of users is essential for future quantum information science and technology applications.
However, beyond simple point-to-point protocols, existing methods struggle with increasingly complex state preparation for a growing number of parties.
Here, we show that, surprisingly, multiparty loophole-free quantum steering, where one party simultaneously steers arbitrarily many spatially separate parties, is achievable by constructing a quantum network from a set of qubits of which only one pair is entangled.
Using these insights, we experimentally demonstrate this type of steering between three parties with the detection loophole closed.
With its modest and fixed entanglement requirements, this work introduces a scalable approach to rigorously verify quantum nonlocality across multiple parties, thus providing a practical tool towards developing the future quantum internet.

\end{abstract}

\maketitle

Quantum nonlocality is a resource for secure communications and distributed information tasks~\cite{Slussarenko2019, Nielsen2000, Uola2020}. The latter include distributed quantum computing~\cite{Cuomo2020}, quantum cryptography \cite{Branciard2012,Xiang2017,Mondal2019}, randomness certification \cite{Law2014,Mattar2017,Joch2022}, quantum state teleportation \cite{Bennett1993,Hermans2022}, and long-range sensor nets such as extended baseline optical telescopes~\cite{Gottesman2012}.

Nonlocal communication protocols prevent eavesdroppers or malicious parties from sabotaging or gaining information from sensitive communication and guarantee unconditional security \cite{Hen15, giustina15, shalm15}.
This is done by having networks of separate parties (usually two) measure correlations and violate a Bell inequality or steering inequality.
Quantum (or Einstein-Podolsky-Rosen) steering~\cite{Wiseman2007, Jones2007} is a form of nonlocality which distinguishes itself from Bell nonlocality in several ways \cite{Cavalcanti2017, Uola2020}. Notably, by trusting a subset of parties, quantum steering is more robust to loss and noise on the untrusted channels~\cite{Ben12, Kocsis2015, Weston2018a}.

Steering is often studied in point-to-point scenarios, but more than two spatially separate parties are required for most networking applications. Establishing multiparty quantum networks requires understanding new and more complex nonlocal and causal relationships beyond the well-studied two-party scenario described by Bell inequalities~\cite{Bell1964}. For quantum steering in a multiparty scenario, many network topologies can arise, as any observer in the network could play the role of a trusted or untrusted party within a given steering task. 
This leads to novel nonlocality phenomena such as network quantum steering~\cite{Jones2021} or collective steering~\cite{He13, Armstrong2015,Kog17, Cai20}, where subsets of parties jointly attempt to steer other subsets.

The case we consider here is multiparty steering scenarios, where an untrusted party can simultaneously demonstrate bipartite steering of multiple trusted parties.

Increasing the number of parties in multiparty steering by directly extending known approaches is a daunting endeavour, as it entails scaling up the quantity or complexity of entangled resources. For example, in Ref.~\cite{Joshi2020}, a scenario where one (untrusted) Alice steers $N$ (trusted) Bobs requires Alice and each of the Bobs to share an entangled pair, requiring $2N$ total qubits. Alternative approaches rely on complex quantum states with entanglement depth larger than $2$~\cite{Guhne2005}, such as $N$-qubit W, GHZ, cluster or graph state~\cite{Cavalcanti2015, Li2015, McCutcheon2016, Lu2020}. 

Multiparty steering with a GHZ state was demonstrated in a proof-of-principle experiment involving two photons and two degrees of freedom in Ref.~\cite{Cavalcanti2015}. However, that experiment did not close the detection loophole. Additionally, two of the three qubits were encoded in one photon's polarization and path, so these qubits could not be spatially separated as required for remote parties. This lack of spatial separation is also seen in other works \cite{Armstrong2015, Li2015}. A different approach involves using a single entangled pair by using sequential weak measurement~\cite{Curchod2017, Choi2020, Zhu2022}. These works require all trusted parties to share a common quantum channel and rely on post-selection.

Here, we overcome the problem of growing entangled resource needs by introducing a scalable and resource-efficient approach to multiparty steering. We discover the simplest possible states---considering the number of qubits, entanglement depth, number of entangled pairs involved, and amount of entanglement---which can demonstrate loophole-free quantum steering where one party steers an arbitrary number $N$ of spatially separated parties simultaneously. Our state is built from a single entangled qubit pair plus $(N-1)$ qubits in a product state. We perform a photonic experiment for $N=2$, demonstrating simultaneous steering between one untrusted party and two trusted parties with the detection loophole closed. Our approach, in principle, allows for the steering of an infinite number of parties if there is no noise or loss. Even with the noise of our currently produced states, we predict steerability of up to $N=26$ trusted parties is possible.

\section{Results}

\subsection{Scenario}

\begin{figure}[htbp]
\centering
\includegraphics[width=1\linewidth]{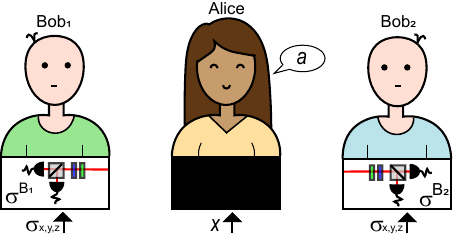}
\caption{\textbf{Multiparty quantum steering scenario.} $N+1$ parties share a quantum state. 
(Untrusted) Alice receives measurement instructions $x$ and reports outcomes $a$. Each (trusted) Bob makes Pauli projections from a tomographically complete set in each protocol run and can independently verify if Alice's measurements can steer their respective particle.}
\label{fig:monogamy_diagram}
\end{figure}

Consider a network involving $N+1$ spatially distant parties, composed of one untrusted Alice and $N$ trusted Bobs, as in Fig.~\ref{fig:monogamy_diagram}. The parties in this network share a quantum state such that each observer has one out of $N+1$ qubits. 
Through a quantum steering test, Alice attempts to convince each of the Bobs simultaneously that she shares nonlocality with them.
The test result is evaluated based on measurement outcomes reported by the parties over repeated protocol runs.

Since Alice is untrusted, we treat her measurement device as a black box, making no assumptions about how her outcomes are generated.
Alice receives classical instructions labelled by $x$, specifying which out of a set of predetermined measurements to perform in each protocol run, and she broadcasts the corresponding outcomes $a\in\{+,-,\emptyset \}$ 
to the other $N$ parties. 
Here, $\emptyset$ represents the null outcome, corresponding to an event where Alice has received a measurement instruction but reports no outcome. The probability of reporting a non-null outcome per protocol run is called Alice's efficiency.
 
In every protocol run, each Bob (labelled $B_n$) can perform a projective measurement on his qubit (from a tomographically complete set). 
Here, no option of a null outcome is required because the Bobs---who are trusted and collectively decide what constitutes a run of the protocol---exclude instances where their measurement devices do not report an outcome. 
Over time, each Bob sorts his measurement statistics by Alice's announced results, thus creating ensembles of locally observed quantum states normalised by the probability of Alice observing a corresponding measurement outcome. These ensembles are commonly known as assemblages \cite{Cavalcanti2017}. Importantly, an assemblage contains all the information relevant to deciding the result of a quantum steering test. 

For a given Alice-Bobs' bipartite state $\rho$, the $n^\mathrm{th}$ Bob's assemblage is a collection of unnormalized quantum states  $\sigma^{B_n}_{a|x} = \Tr_A \left[ (E_{a|x} \otimes I )\rho_{A, B_n} \right]$. Here, $\rho_{A, B_n}\coloneqq \Tr_{\neg A, B_n}[\rho]$ is the reduced system shared by Alice and Bob $n$, while $\{E_{a|x}\}_a$ is the POVM (over $a$) for each of Alice's settings $x$.
Any bona fide assemblage must contain only positive semi-definite matrices which satisfy the no-signalling condition, $\forall x, x',\sum_a \sigma^{B_n}_{a|x} = \sum_{a}\sigma^{B_n}_{a|x'} \eqqcolon \rho_{B_n}$.
Here, we employ a convex optimization technique to reconstruct the assemblages (see Methods), which ensures these properties.

Alice has \emph{steered} the $n^\mathrm{th}$ Bob if every element of the assemblage cannot be written as a coarse-graining over local-hidden-states $\{ \rho^{B_n}_\lambda\}_\lambda$ detected by the $n^\mathrm{th}$ Bob \cite{Wiseman2007}. That is, there does \textbf{not} exist a set of normalized states $\{\rho_\lambda^{B_n}\}_\lambda$, and probability distributions $p_\lambda$ and $P_\lambda(a|x)$ over $\lambda$ and $a$ respectively such that,
\begin{equation}
\forall a,x,\sigma^{B_n}_{a|x} = \sum\limits_\lambda p_\lambda P_\lambda(a|x) \rho^{B_n}_\lambda.
\end{equation}

If Alice’s efficiency is below unity, a common approach
is to postselect by ignoring the null outcomes and introducing the fair sampling assumption that Alice’s reported measurement outcomes accurately represent the
total statistical sample. However, as in any rigorous nonlocality test, such experimental assumptions open loopholes allowing false nonlocality verification.

The loophole associated with the fair sampling assumption is the detection loophole; it allows Alice to
cheat by not reporting some of her measurement outcomes to mimic a steerable assemblage. Alice's null outcome instances need to be taken into account to close
this loophole and drop the need for the fair sampling assumption. This amounts to placing a lower bound on
Alice’s efficiency, which she must surpass to steer Bob
via her non-null outcomes.

We implement this using inequalities of the following form, whose satisfaction certifies detection-loophole-free steering~\cite{Bak22}:
\begin{equation}
\epsilon^{\star}(\{\sigma_{a|x}^{B_n}\}) \le \epsilon_{\text{exp}}.
\label{eq:main_steeringinequality}
\end{equation}
Here, $\epsilon_{\text{exp}}$ is Alice's measured efficiency --- the proportion of her non-null results --- in the experiment, while $\epsilon^{\star}$ is a cutoff efficiency. This is a function of the assemblage $\{\sigma_{a|x}^{B_n}\}_{a|x}$  observed by the $n^\mathrm{th}$ Bob when Alice announces any non-null outcome, i.e. $a\neq \emptyset$.  The cutoff efficiency $\epsilon^{\star}$ is the maximum efficiency that could allow a cheating Alice to exploit the fair sampling assumption.
This cutoff efficiency is determined through a semi-definite program (SDP) (see Methods).

\subsection{Multiparty, loophole-free steering with a resource-efficient state}

A crucial step towards realising loophole-free multiparty steering demonstrations lies in preparing an appropriate multiparty entangled quantum state. Here, we introduce a practical quantum state that can demonstrate loophole-free steering from Alice to each of the Bobs independently, based on a single entangled pair of qubits and $N-1$ pure single qubits.

The mixed global quantum state we prepare consists of $N+1$ qubits, acting on a composite Hilbert space $\mathcal{H}_A \otimes \mathcal{H}_{B_1} \otimes \cdots \otimes \mathcal{H}_{B_N}$.
Below, we omit system labels when there is no risk of confusion.
Consider the family of two-qubit pure states $\ket{\psi_\alpha}:=\sqrt{\alpha}\ket{11} + \sqrt{1-\alpha}\ket{00}$, that is entangled for $\alpha\in(0,1)$.
From $\ket{\psi_\alpha}$, we construct the family of states between the $N+1$ parties
\begin{equation}
\rho_{\alpha, N} := \frac{1}{N}\sum\limits_{n=1}^{N} {\mathbf{V}}_{1,n} \left(\ketbra{\psi_\alpha}\otimes\ketbra{0}^{\otimes N-1} \right),
\label{eq:mainclass}
\end{equation}
where ${\mathbf{V}}_{m,n}$ is a superoperator that swaps subsystems $B_m$ and $B_n$: $\mathbf{V}_{m,n}(\rho_{B_m}\otimes\rho_{B_n})\equiv\rho_{B_n}\otimes\rho_{B_m}$.
As we show later, this state permits loophole-free quantum steering with an arbitrary number of parties and can be prepared using deterministic quantum gates.  

For now, we concentrate on the case of $N=2$, where we obtain the 3-party state
\begin{multline}
\label{eq:mainclass3}
    \rho_{\alpha,2} \coloneqq \frac{1}{2} \left(\ketbra{\psi_\alpha}_{A,B_1}\otimes\ketbra{0}_{B_2} \right. \\
    \left. +\ketbra{\psi_\alpha}_{A,B_2}\otimes\ketbra{0}_{B_1} \right),
\end{multline}
distributed between the untrusted party Alice and the trusted parties Bob 1 and Bob 2, as indicated by the system labels. 
The state represents a mixture of two cases in which one part of $\ket{\psi_\alpha}$ remains at Alice's station, and the other is sent to either Bob 1 or Bob 2, with the remaining Bob receiving a single photon in the computational zero state. 
The cutoff efficiency $\epsilon^{\star}(\{\sigma_{a|x}^{B_n}\})$ of the assemblage produced for Bob 1 and Bob 2, assuming Alice performs three dichotomic measurements corresponding to Pauli operators, is illustrated as the purple solid line in Fig.~\ref{fig:main}.
Quantum steering is in principle possible whenever $\epsilon^{\star}(\{\sigma_{a|x}^{B_n}\})$ is below one. This occurs for  $\alpha \in(0,2/3)$ and the minimal value of $\epsilon^\star$ is achieved in the singular limit $\alpha\rightarrow0$, i.e.~at vanishing entanglement of $\ket{\psi_\alpha}$ (as measured e.g.~by concurrence). 

When the family of states of Eq.~(\ref{eq:mainclass3}) is modified slightly through the addition of noise, the singularity disappears, as shown by the purple \textit{dashed} line of Fig.~\ref{fig:main}. Although counter-intuitive, this finding that a small value of $\alpha$ is optimal aligns well with the fact that for the case of two parties, pure states with a small amount of entanglement exhibit detection-loophole-free nonlocality at a lower efficiency bound than a maximally entangled state in the cases of both Bell nonlocality~\cite{clauser74} and quantum steering~\cite{vallone_steering} tests.

\subsection{Experimental implementation of three-party, detection-loophole-free steering}
\begin{figure}[htbp]
\centering
\includegraphics[width=1\linewidth]{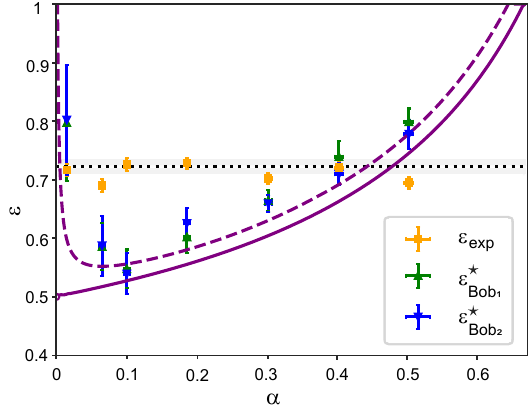}
\caption{\textbf{Data demonstrating detection-loophole-free steering as a function of the quantum state parameter \bm{$\alpha.$}} Bipartite quantum steering takes place when $\epsilon_{\textrm{exp}}>\epsilon^\star_{BobN}$. The green (blue) points illustrate the minimum efficiency required to demonstrate steering for the group Alice-Bob 1 (Alice-Bob 2). The orange squares correspond to Alice's minimum efficiency across her measurement settings, and the black dotted line is the mean across the different $\alpha$ values, with the grey-shaded region illustrating $\pm$ one standard deviation across all measured efficiencies. All data points include horizontal error bars, though frequently smaller than the data marker.
The purple solid line depicts the theoretical lower bound for the cutoff efficiency, $\epsilon^\star(\alpha,2)$ in Eq.~(\ref{eq:3settings_analytic}), for the ideal target family of states from Eq.~(\ref{eq:mainclass3}). 
The purple \textit{dashed} line illustrates numerically determined cut-off efficiencies $\epsilon^\star$ (see Methods) for a more realistic family of states, consistent with experimental tomographic reconstructions of the target states before the multiparty steering test; these are well-modelled by the action of the depolarizing channel on each target, $\Delta_\eta(\rho) = \eta\rho+(1-\eta)I/4$ with $\eta=0.9931$.}
\label{fig:main}
\end{figure}

\begin{figure*}[t]
\centering
\includegraphics[width=1\linewidth]{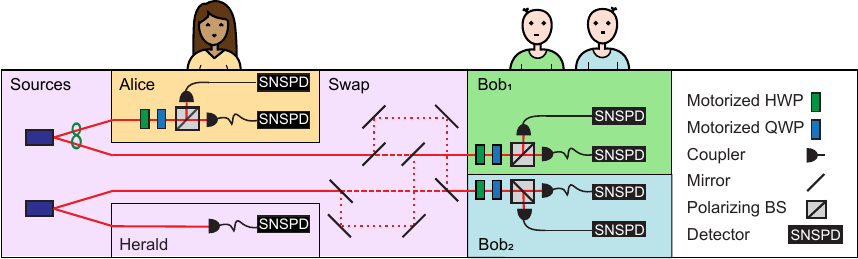}
\caption{\textbf{Experimental layout.} Sources: the top photon source produces the entangled state $\ket{\psi_\alpha}$ from Eq. (~\ref{eq:mainclass3}) and distributes half to Alice and the other to the Swap stage. The bottom source produces the unentangled $\ket{0}$, sending one photon to the Swap stage and the other to the heralding detector. Swap: mirror mounted on a linear translation stage displace the beam's trajectory with ~0.5 probability, producing the state's mixture. While the swap operation occurs (mirror positions are shown in the figure), the light follows the dotted lines; otherwise, the mirrors are removed from the beam path, and the light will follow the dashed line; path segments common in both cases are solid lines.  Alice, Bob 1, and Bob 2: each party perform projective measurements with motorised wave plates and a polarizing beam splitter. Coincidence are measured with superconducting nanowire single-photon detectors (SNSPD). See Methods for more details}
\label{fig:exp}
\end{figure*}

We implement the protocol using photons as the experimental platform, encoding qubits in the polarisation degree of freedom of single photons. Our experimental setup is shown in Fig.~\ref{fig:exp}. To prepare the state of Eq.~(\ref{eq:mainclass3}), we first generate an entangled photon pair in the state $\ket{\psi_\alpha}$ and a heralded single photon using two high-efficiency, telecom-wavelength photon pair sources based on  group-velocity-matched spontaneous parametric down-conversion with a pulsed pump~\cite{Tischler2018}. One half of the entangled pair is sent to Alice, and the other half is probabilistically distributed to one of the trusted parties, Bob 1 or Bob 2, with the other Bob receiving the heralded single photon. Alice is instructed to perform one out of a set of three projective measurements on her photon. The Bobs each perform a quantum state tomographic measurement on their respective photon.

Bob 1 and Bob 2 do not need to share a quantum channel but require the other party to indicate through a classical channel if they obtain an outcome. In the cases where both Bobs announce that they have obtained
an outcome, the run goes ahead, regardless of whether Alice announces a measurement outcome or claims to have lost her photon. 

To build up statistics, the protocol is repeated for each of Alice's measurement settings and various tomographic projections for the Bobs. Afterwards, Bob 1 and Bob 2 can reconstruct and analyze their assemblages (see Methods for details), determine Alice's efficiency based on the proportion of runs she reported an outcome, and test the inequality of Eq.~(\ref{eq:main_steeringinequality}). Alice's efficiency includes all losses associated with her photon, from state preparation through to her detection efficiency. Further experimental details are provided in Methods.

The results of steering tests for different states are shown as the data points in Fig.~\ref{fig:main}. Multiparty steering is demonstrated when $\epsilon_{\textrm{exp}} > \epsilon^{\star}_{Bob1}$ and $\epsilon_{\textrm{exp}} > \epsilon^{\star}_{Bob2}$, which is clearly observed at $\alpha$ values between $0.065$ and $0.3$. The most statistically significant steering occurs at $\alpha \approx 0.1$, where we measured a minimum efficiency 5.34 (5.35) standard deviations above the cut-off for Bob 1 (Bob 2). This is thanks to our high experimental heralding efficiencies above 0.69.   
Since the cut-off efficiencies are numerically determined from the experimentally reconstructed assemblages, the steering demonstrations are conclusive independent of how well the experimental states approximate the target states of Eq.~(\ref{eq:mainclass3}).

\subsection{Extension to more parties}
\label{Extendibility}

As previously mentioned, our steering scenario can be extended to $N+1$ parties by considering the generalised state $\rho_{\alpha,N}$ in Eq.~(\ref{eq:mainclass}). 

\begin{figure}[htbp]
\begin{center}
\includegraphics[width=1\linewidth]{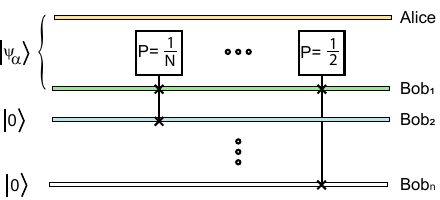}
\caption{\textbf{Quantum circuit diagram for the construction of the global \bm{$N+1$} qubit state Eq. (\ref{eq:mainclass}).}  
One half of the entangled pair $\ket{\psi_\alpha}$ is sent to Alice, and the other is randomly sent to one of the $N$ Bobs through a sequence of deterministically implementable random-swap gates. }
\label{fig:circuit_diagram}
\end{center}
\end{figure}

A method for creating the states is illustrated in Fig.~\eqref{fig:circuit_diagram}, where a sequence of $N-1$ deterministically implementable gates successively acts on one-half of the entangled pair and one of the ancilla qubits, which are initialized to $\ket{0}$. 
Each gate is a random-swap gate, a completely-positive trace-preserving map between states acting on the Hilbert spaces of $B_1$  and $B_n$, $\Phi_n \coloneqq p_n {\mathbf{V}}_{1,n} + (1-p_n) \mathbf{I}_{1,n}$.
Here $\mathbf{I}_n$ is the identity superoperator, and $p_n \coloneqq 1/(N-n+1)$ is the swapping probability for the gate.
The ordered composition of these gates gives the desired output of the circuit, 
\begin{equation}
\rho_{\alpha, N} \equiv \Phi_{N-1} \circ \dots \circ \Phi_n \circ \dots \circ \Phi_1 \left(\ketbra{\psi_\alpha}\otimes\ketbra{0}^{\otimes N-1} \right).
\end{equation}
This state can be created in a resource-efficient way and with several appealing properties.
\begin{itemize} 
\item \emph{Number of qubits:} Each party only requires one qubit, unlike the simple extension of two-party steering where an entangled pair of qubits gets distributed between Alice and each of the Bobs, which would involve $2N$ qubits. 
\item \emph{2-Producibility:}
Multiparty quantum states can be characterised using the concept of $k$-producibility~\cite{Guhne2005}.
A pure quantum state $\ket{\psi}$ is $k$-producible if it can be written as a composition of quantum states involving at most $k$ parties.
That is, $\ket{\Psi} = \ket{\psi_1}\otimes\ket{\psi_2}\otimes\ket{\psi_3}\otimes \dots$ where each $\ket{\psi_i}$ is a state shared between $k$ parties.
Similarly, a mixed state is $k$-producible if it can be decomposed as a mixture of $k$-producible pure states. If a state is not ($k-1$)-producible, but \emph{is} $k$-producible, then its entanglement depth is $k$. The states from Eq.~(\ref{eq:mainclass}) are 2-producible and have entanglement depth 2, independently of the number of parties involved---unlike $N+1$-party GHZ and W states which both have entanglement depth $N+1$.
\item \emph{Number of entangled pairs:} The state preparation only requires a single pair of entangled qubits, which is a stronger constraint on the required resources than 2-producibility alone.
\item \emph{Deterministic implementation of gates:} The creation of the states involves gates that can be implemented on photons deterministically, in the sense that the gates do not require postselection or heralding, unlike controlled-NOT gates. 
\end{itemize}

In Section IV. of the Supplementary Information, we derive an exact expression for $\epsilon^{\star}(\{\sigma_{a|x}^{B_n}\})$ of 
the assemblage prepared for each Bob, in the case when Alice performs the three dichotomic Pauli measurements.
The result is
\begin{widetext}
\begin{equation}
\epsilon^{\star}(\alpha,N) = \frac{2N -\alpha  (N+1)+2 \sqrt{2(1-\alpha)} \sqrt{N-\alpha }+\sqrt{\alpha 
   (N-1) \left( (\alpha +4) N -5 \alpha -4 \sqrt{2(1- \alpha) }
   \sqrt{N-\alpha }\right)}}{2 \left(2
   \sqrt{2(1-\alpha)} \sqrt{N-\alpha }+N+2-\alpha -2 \alpha  N\right)}.
\label{eq:3settings_analytic}
\end{equation}
\end{widetext}

Detection-loophole-free steering could be observed when this cutoff efficiency is strictly below unity, which occurs for $\alpha \in(0,2/(N+1))$. From a theoretical point of view, this bound on efficiency has various interesting properties.
The minimal value of $\epsilon^\star$ is achieved in the singular limit $\alpha\rightarrow0$,
 the same limit in which the concurrence of $\ket{\psi_\alpha}$ vanishes.
This generalises the earlier observation from the $N=2$ case (purple solid curve in Fig.~\ref{fig:main}) in the idealised scenario with exact measurements and no noise. Here, the required detection efficiency is only $\epsilon^\star = (1+\sqrt{2/N})^{-1}$, which is remarkable for a state produced from a single pair of entangled qubits.
\begin{figure}[htbp]
\centering
\includegraphics[width=\linewidth]{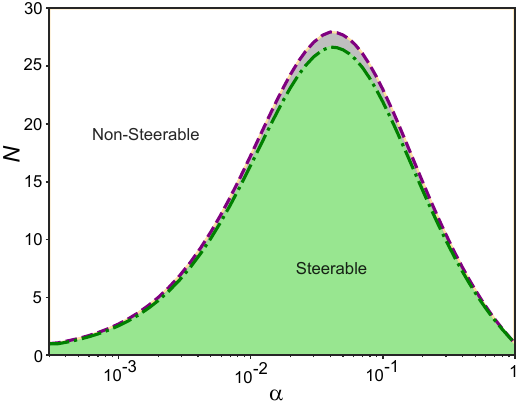}
\caption{
\textbf{The limit to the number of Bobs \bm{$N$} which could be steered, using noisy network states.} This result is based on the
two-qubit state used in the experiment, Eq. (\ref{eq:mainclass3}), with the presence of experimental noise, which is estimated as per the caption of Fig.~\ref{fig:main}. 
The simulation here assumes conditions that are, apart from the noise, ideal: unit detection efficiency and infinite measurement settings. The numerically uncertain region is shaded in grey.}
\label{fig:extendibility}
\end{figure}

The analytic bound in Eq.~\eqref{eq:3settings_analytic} demonstrates that there always exists a value of Alice's efficiency which permits arbitrarily many Bobs to be steered in a loophole-free way.
However, this assumes an idealised scenario where noise is absent from the experiment. We thus also investigate the scalability of steering for the states defined by Eq.~\eqref{eq:mainclass}, but with the addition of noise at the level we observed in our experiment. 
We implement the numerical program from~\cite{Ngu19} to certify steerability from Alice to each Bob under all qubit projective measurements, that is, in the limit of an infinite number of measurement settings.
The results of these simulations are illustrated in Fig.~\ref{fig:extendibility}.
Remarkably, there exists a judicious choice of $\alpha$ such that up to $26$ Bobs could be steered simultaneously.

\section{Discussion}
In this work, we demonstrate a multiparty quantum nonlocality experiment with spatially separated parties in the discrete variable regime -- the first such experiment to close the detection loophole. We achieve this by a 
novel, resource-efficient state preparation scheme that allows multiparty quantum steering. Remarkably, using just a single pair of entangled qubits, one party can, in theory, steer arbitrarily many other parties. 
Importantly, this approach is robust to the commonly overlooked effects of photon loss and noise: our steering inequality takes these effects into account and can be satisfied under demanding but nevertheless viable conditions.
We apply our method to experimentally satisfy a detection-loophole-free quantum steering inequality in a network of three spatially separated parties by $5.35$ standard deviations. We thereby demonstrate multiparty steering where one untrusted party simultaneously steers two trusted parties.
We show steering of up to $N=26$ is possible with realistic experimental quantum state fidelities and ideal efficiencies.
Unlike other methods, our approach is scalable as it does not require  heralded or postselected gates to generate the steerable state, and neither does it require increasing entanglement for larger numbers of parties. Our protocol does not rely on sequential measurements, such as those used in continuous variable protocols. Thus, we do not require a quantum channel connecting our  trusted parties. In the experiment, we focused on closing the detection loophole, which is crucial for real-world implementations.

Verification of quantum nonlocality is essential for the implementation of secure quantum networks. For the experimental design and data analysis, we use semi-definite programming to determine steering bounds and novel techniques to reconstruct quantum assemblages using maximum-likelihood estimation; these can also find wider applications in other quantum steering contexts. 

Future directions include an implementation of a fully loophole-free protocol, implementing a larger network with more parties, and including a larger number of untrusted elements in various topologies.

Our work demonstrates a realistic method for steering verification in a large-scale quantum network. We show how steering-based quantum networks of tens of users can be implemented. From a secure quantum communication application side, our protocol allows a trusted network to introduce an untrusted member, which may be useful in user authentication, such as banking, multi-factor authentication, and implementations of a quantum internet.

\section{Methods}

\subsection{Detection-loophole-free steering and measurement design}

    Since imperfections such as noise are experimentally unavoidable and lead to deviations from the ideal behaviour, we follow a multi-step process to adapt Alice's measurement directions and determine the correct cutoff efficiency $\epsilon^{\star}$. In the design stage prior to the steering test, we aim to prepare seven members of the ideal family of two-qubit states $|\Psi_{\alpha}\rangle$, and perform quantum state tomography to characterise the level of noise. We find the average fidelity between the prepared and ideal states to be 0.987 $\pm${0.003}. 
    Based on this estimate, we perform a global differential evolution optimisation routine to find Alice's measurement directions that minimise the $\epsilon^{\star}$ we expect to obtain in the steering test. 
    We then perform the steering test using these measurement settings, which provide the assemblages and Alice's efficiency $\epsilon_{\mathrm{exp}}$. 
    The assemblages are used to determine $\epsilon^{\star}$ via a semi-definite program, and the outcomes of the steering test are decided based on Eq.(\ref{eq:main_steeringinequality}),
    $\epsilon^{\star}(\{\sigma_{a|x}^{B_n}\}) \le \epsilon_{\text{exp}}$.
    The left side of this inequality is the maximum efficiency, i.e.\ average proportion of non-null results, that Alice could have announced in an experiment preparing the assemblages $\{\sigma_{a|x}^{B_n}\}$. 
    This quantity is found by solving the SDP~\cite{Bak22}:
    \begin{equation}
    \begin{aligned}
    & \underset{\{\sigma_\lambda\}_\lambda}{\text{max}} & & \epsilon &\\
    & \text{s. t.} & & \sum_\lambda D(a |x,\lambda)\sigma_\lambda = \epsilon\sigma_{a|x}^{B_n} & \forall x, a\neq\emptyset\\
    & & & \sum_\lambda \sigma_\lambda = \rho_{B_n} & \\ 
    & & & \sigma_\lambda \geq 0 & \forall \lambda. \\
    \end{aligned}
    \label{eq:LTSDP}
    \end{equation}
    Here, $\{ D(a |x,\lambda) \}_\lambda$ is the set of deterministic probability distributions mapping $x$ to \emph{all} outcomes (both null and non-null).
    This optimization problem can be interpreted as a functional that maps from the space of  assemblages to real numbers in $(0,1]$.
    For a given input assemblage, exact values for $\epsilon^{\star}$ can be found numerically.
    
    When the assemblage is prepared by Alice measuring on her part of $\rho_{\alpha, N}$, the assemblage (and hence $\epsilon^\star$) will depend on the measurements she chooses to make. 
    Assuming these are dichotomic projective measurements $\{\Pi_{a|x} \}_{a,x}$ on her qubit, 
    we decompose them as $\Pi_{a|x} = (I + a \bm{r}_x\cdot\bm{\sigma})/2$, with $\bm{r}_x \coloneqq (\sin\theta_x \cos\phi_x, \sin\theta_x\sin\phi_x, \cos\theta_x)^T$.
    We then wish to find Alice's minimum detection efficiency by searching over these measurement settings.
    That is, for various values of $\alpha$, we seek solutions to
    \begin{equation}
    \min\limits_{\{\Pi_{a|x}\}_{a,x}} \epsilon^{\star}(\{\Tr_A \left[ (\Pi_{a|x} \otimes I ) \tilde{\rho}_{\alpha,2}\right]\}).
    \end{equation}
    The quantum state $\tilde{\rho}_{\alpha,2}$ appearing in the objective function is experimentally determined from averaging over quantum state tomographies to account for noise. 
    We perform this outer search over the measurement settings heuristically, using a global differential evolution optimization routine.
    The results of these searches are summarised in Supplementary Information II.

\subsection{Assemblage tomography}

\begin{figure}[htbp]
\includegraphics{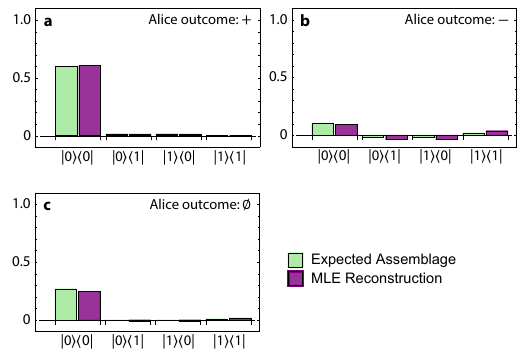}
\caption{\textbf{The actual reconstructed assemblage tomography result versus the expected experimental assemblage for the largest steering demonstration \bm{$\alpha=0.1$}.} {\bf a}, {\bf b}, and {\bf c}  show the Alice-Bob 1 assemblage with the first group of POVM setting (see Supplementary Information III) corresponding to Alice's three outcomes: $+$, $-$, and $\emptyset$, respectively. Green (purple) bars correspond to the expected (reconstructed) assemblage. Imaginary components are not plotted as they are all $<0.002$.}

\label{fig:ass_exam}
\end{figure}

Here we formulate our task as finding the most likely assemblages and Bobs' states to generate the given experimental data:
\begin{equation}
\begin{aligned}
     \underset{\{\sigma_{a|x}\}_{a,x}, \rho_{B_n}}{\text{max}}  & \sum_{a,b,x,y}C(b|y)|_{a|x}\log \Tr (E_{b|y}\sigma_{a|x}) \\
     \text{s. t.\ \ } & \forall x, \ \ 
\sigma_{\emptyset|x}=(1-\epsilon)\rho_{B_n},\\
     & \forall x, \ \ \sum_{a}\sigma_{a|x} = \rho_{B_n} ,\\
     &  \Tr\rho_{B_n}=1  ,\\
     & \forall a,x, \ \  \sigma_{a|x} \geq 0, 
\end{aligned}\label{Prob_Assem}
\end{equation}
where $C(b|y)|_{a|x}$ is the experimental count of outcome $b$ from POVM $E_{b|y}$, and $\epsilon$ is calculated from $\sum_{x,y,b; a\neq \emptyset}C(b|y)|_{a|x}/\sum_{x,y,a,b}C(b|y)|_{a|x}$ before solving Problem (\ref{Prob_Assem}). Every quantity in (\ref{Prob_Assem}) except the indices is indexed by $\alpha$ and $B_n$, which we mostly omit to ease the burden of expression. 
We solve (\ref{Prob_Assem}) for each $\alpha$ and $n$ independently, maximizing the logarithm of the likelihood of the unknowns generating the data in experiment. 
For fixed $\alpha$ and $n$, the maximum likelihood estimation (MLE) problem in (\ref{Prob_Assem}) is a conic optimization problem \cite{Boyd04}, which is more general than  an  SDP, and can be solved using standard software.
Here we employ the  YALMIP toolbox \cite{yalmip} and call the MOSEK solver \cite{mosek} under a MATLAB environment, and a selection of tomography results from experiment data is shown in Fig. \ref{fig:ass_exam}.

We then perform hypothesis testing to check the degree to which our tomography results are consistent with the experimental data (see Supplementary Information V). We obtain a test value of $572.8553$ from our tomography results, smaller than the threshold of the critical region $595.1683$ (corresponding to a significance level $s_\alpha=5\%$), indicating an acceptance of the null hypothesis that the MLE result matches the true value of the assemblage..

The tomography for $\alpha=0.015$ was omitted from this testing as this data point has a large statistical uncertainty arising from diminishing photon counts as $\alpha$ decreases, which causes the hypothesis testing to fail.

\subsection{Experimental details}

The experiment, Fig.~\ref{fig:main}, uses two sources of single photon pairs based on the design of~\cite{Tischler2018}. One source provides a tunable entangled two-qubit state $|\Psi_\alpha\rangle$ and the other the single qubit state $\ket{0}$ (with the fourth photon detected to herald the photon that encodes the single qubit). We use a 775nm wavelength Ti:sapph pump laser with 1ps pulse length and 80 MHz repetition rate. The pump is split on a 50:50 beam splitter. The resulting beams are used to pump two 15 mm long periodically poled potassium titanyl phosphate crystals with approximately 100mW pump power per crystal, each resulting in degenerate type-II spontaneous parametric down-conversion (SPDC) at telecom wavelength.

In each protocol run, the entangled state $|\Psi_\alpha \rangle$ is shared between Alice and one other trusted party, and the third party receives the $\ket{0}$ state. For each quantum state with a different value of $\alpha$, approximately $10^{5}$ runs are performed---a run being an event where both trusted parties and the heralding detector detect a photon. Of those runs, in approximately half $(0.53\pm{0.05})$ of the cases, Bob 1 receives part of the entangled state, and in the other cases, Bob 2 does. The swap operation, which determines whether Bob 1 or Bob 2 receives half of the entangled state, is implemented by a linear translation stage that moves mirrors into the beam paths, redirecting their spatial modes. 

Each of the three parties can perform projective measurements using automated quarter-wave and half-wave plates, a polarizing beam splitter, and two superconducting nanowire single-photon detectors (SNSPD). Alice's detectors, the only detectors whose efficiency directly impacts the protocol's success, have an efficiency of $\sim 90\%$.

The data for the experiment are recorded in 720-second batches. A batch consists of the linear translation stage remaining in one position while Alice and the Bobs sequentially perform their measurements in a predetermined order. We then average the detection efficiency of our SNSPDs by repeating these measurements with the wave plates rotated to swap which detector receives which outcome of the POVMs. Alice's measurement settings come from an optimisation routine (see Supplementary Information II), and the Bobs perform tomographic measurements before moving on to the next combination of measurement settings. After each combination, the linear translation stage shifts  (or does not shift) and measurement iterations begin again.
The data files for each batch are summed to obtain the mixture, then organised into outcome groups between Alice and each of the Bobs, and we extract the heralding efficiencies ($\epsilon_{a,x}$). We reconstruct the assemblages through conic optimization. From these assemblages, we use a SDP Eq.~(\ref{eq:LTSDP}) to calculate the cutoff efficiency for the trusted parties, $\epsilon_{{\rm Bob}_1}^{\star}$ and $\epsilon_{{\rm Bob}_2}^{\star}$.

\section{Data availability}

The data supporting the plots within this paper and other study findings are available in the Supplementary information. Further data is available from corresponding authors upon reasonable request.

\section{Code availability}
Details are available from the authors upon reasonable request.

\section{Acknowledgements}
This work was supported by the Australian Research Council Centre of Excellence CE170100012 and the National Natural Science Foundation of China (No. 12288201). A.P., T.J.B., and Q.C.S. acknowledge support from the Australian Government Research Training Program (RTP).
We acknowledge the support of the Griffith University eResearch Service \& Specialised Platforms Team and the use of the High-Performance Computing Cluster ``Gowonda'' to complete this research.

\section{Contributions}
A.P. designed and implemented the experiment and analysed the data with the guidance of S.S. N.T. and G.J.P. Y.W. designed and coded the assemblage tomography and performed the hypothesis testing. T.J.B. and H.M.W. conceptualized the states and theoretically investigated how they can demonstrate steering. Q.C.S., T.J.B., and H.M.W. analysed scalability. L.K.S., V.B.V., and S.W.N. developed the high-efficiency SNSPDs. All authors discussed the results and contributed to this manuscript. Correspondence and material requests should be addressed to A.P., T.J.B., N.T., and H.M.W.

\appendix

\clearpage
\pagebreak
\widetext
\begin{center}
\textbf{\large Supplementary Information\\ for \\ Scalable multiparty steering based on a single pair of entangled qubits}
\end{center}

\vspace{20pt}

\setcounter{equation}{0}
\setcounter{figure}{0}
\setcounter{table}{0}
\setcounter{page}{1}
\makeatletter
\renewcommand{\theequation}{S\arabic{equation}}
\renewcommand{\thefigure}{S\arabic{figure}}
\renewcommand{\bibnumfmt}[1]{[S#1]}
\renewcommand{\citenumfont}[1]{S#1}

\setcounter{secnumdepth}{2}

\section{Data demonstrating detection-loophole-free steering}
\begin{table}[htp]

\label{app:steeringdata}
\caption{\textbf{Data demonstrating detection-loophole-free steering.} $\epsilon_{mean}$ is the average value of $\epsilon{exp}$ across all measurements for a given $\alpha$. Reported uncertainties correspond to one standard deviation of 200 Monte Carlo simulations adding Poisson noise to the raw data and repeating the MLE and steering test.}
\centering
    \begin{tabular}{r|c  | c  | c  | c  | c | c}
     $\alpha$ & $\epsilon_{\rm exp}$ & $\epsilon_{\rm mean}$&
      $\epsilon^\star_{\rm Bob_1}$ & $\epsilon^\star_{\rm Bob_2}$ &
      $\epsilon_{\rm ideal(sim)}$ & $\epsilon_{\rm noisy(sim)}$\\
     \hline
0.015& 0.7171(0.0100)&	0.7276(0.0056)&0.7975(0.0993)&	0.8023(0.0934)&		0.5040&	0.5978\\
0.065& 0.6899(0.0116)&	0.7059(0.0062)&	0.5858(0.0402)&	0.5873(0.0513)&	0.5171&	0.5521\\
0.100& 0.7259(0.0108)&	0.7390(0.0058)&	0.5476(0.0333)&	0.5394(0.0349)&	0.5277&	0.5561\\
0.185& 0.7282(0.0090)&	0.7317(0.0049)&	0.6021(0.0272)&	0.6257(0.0258)&	0.5553&	0.5804\\
0.300& 0.7020(0.0086)&	0.7160(0.0046)&0.6644(0.0182)&	0.6591(0.0145)&		0.6053&	0.6303\\
0.400& 0.7206(0.0095)&	0.7325(0.0051)&	0.7393(0.0273)&	0.7095(0.0186)&	0.6646&	0.6926\\
0.500& 0.6949(0.0099)&	0.7079(0.0047)&	0.7990(0.0227)&	0.7780(0.0247)&	0.7491&	0.7807\\
    \end{tabular}
\end{table}

\section{Alice's optimal measurement angles}
\label{table:AliceAngles}
\begin{table}[htp]
    \caption{\textbf{Alice's numerically found optimal measurement directions.} Three measurements are used for each {$\alpha$} in {$\ket{\Psi_\alpha}$}. Each direction is expressed as a pair of Euler angles (degrees) for the Bloch sphere: the angle from the azimuth $(\theta_n)$ and the zenith $(\phi_n)$.}
    \begin{center}
    \begin{tabular}{r|c c |c c |c c}
    $\alpha$ & $\theta_1$ & $\phi_1$ & $\theta_2$ & $\phi_2$ & $\theta_3$ & $\phi_3$\\ 
    \hline
    0.015 & 0 & 57.2955 & 0.85988 & -3.3977 & -88.0451 & 0.7153\\
    0.065 & 0 & 4.2398 & 84.1949 & -48.6642 & -88.781 & 48.0435\\
    0.100 & 0 & 39.2478 & -60.7293 & 39.2834 & 57.8822 & -41.0124\\
    0.185 & 0 & -33.2033 & -70.5399 & -42.0172 & 64.8311 & 44.2747\\
    0.300 & 0 & -1.1167 & -89.9633 & -2.5873 & 32.5804 & 57.2852\\
    0.400 & 0 & -57.1865 & -53.4757 & -9.8684 & 36.7915 & 10.1887\\
    0.500 & 0 & -13.0359 & 8.8413 & 57.2252 & 89.8365 & 0.50483\\
    \end{tabular}
    \end{center}
\end{table}%

\section{Reconstructed Assemblages}\label{app:assembtomo}
The expected assemblage (left) versus the MLE reconstructed assemblage (right) tomography results for $\alpha=0.1$. Imaginary components are not displayed as they are all $<0.002$.
\begin{table}[!htb]
    \begin{minipage}{.4\linewidth}
      \caption{Expected Assemblage}
      \centering
        \begin{tabular}{c|cccc}
            Outcome&$\ketbra{H}{H}$ & $\ketbra{H}{V}$ &$\ketbra{V}{H}$ &$\ketbra{V}{V}$\\
                \hline
    +&0.6017& 0.0166& 0.0166& 0.0017\\
    -&0.1032& -0.0166& -0.0166& 0.0183\\
    null&0.2676& 0& 0& 0.0076\\
        \end{tabular}
    \end{minipage}%
    \begin{minipage}{.4\linewidth}
      \centering
        \caption{MLE Reconstruction}
        \begin{tabular}{c|cccc}
            Outcome&$\ketbra{H}{H}$ & $\ketbra{H}{V}$ &$\ketbra{V}{H}$ &$\ketbra{V}{V}$\\ 
            \hline
            +&0.6073& 0.0121& 0.0121& 0.0027\\
            -&0.0911& -0.0370& -0.0370& 0.0388\\
            null& 0.2454& -0.0088& -0.0088& 0.0146
        \end{tabular}
    \end{minipage} 
\end{table}

\section{Derivation of Eq.~(6)}. %\eqref{eq:3settings_analytic}.}
\label{app:SI_bound_derivation}

The reduced state between Alice and the $n^\mathrm{th}$ Bob, ($n=1,\dots,N$) from Eq.~\eqref{eq:mainclass} of the main text is
\begin{equation}
\rho^{AB_n}_{\alpha, N} \coloneqq \Tr_{\neg AB_n}\left[\rho_{\alpha, N}\right] = \frac{1}{N} \left( \ketbra{\psi_\alpha} + (N-1)\rho_\alpha\otimes \ketbra{0} \right).
\label{eq:SI_main_two_party_reduce}
\end{equation}
Here, we show the bound in Eq.~(6) of the main text pertains to the assemblage that arises when Alice measures three projective measurements corresponding to the Pauli observables.
Recall from the methods section that the effects of these measurements can be represented by $\Pi_{\pm|x} = (I \pm \bm{\sigma}_{w(x)})/2$, where $w(0)=x, w(1)=y$ and $w(2)=z$.

By symmetry, each Bob is steered to the same assemblage, and so we omit the labels $B_n$ below.
For consistency, below we refer to measurements corresponding to the observables $\sigma_z, \sigma_x$ and $\sigma_y$ according to the ordered index $x=0,1,2$, in that order.
The assemblage detected by each of the $N$ Bobs can therefore be defined as follows.

The ensembles conditioned for the first two measurements are defined by the positive operators
\begin{align}
\sigma_{+1|0}(\alpha,N)&= (1-\alpha)\ketbra{0}, \label{eq:app_assemblage_explicit_first} \\
\sigma_{-1|0}(\alpha,N) &= \frac{\alpha}{2}\left( I + \left( 1-\frac{2}{N} \right)\sigma_z \right), \\
\sigma_{\pm1|1}(\alpha,N) &= \frac{1}{4}\left( I \pm \frac{2}{N}\sqrt{\alpha(1-\alpha)}\sigma_x + \left( 1-\frac{2\alpha}{N} \right)\sigma_z \right), \label{eq:app_assemblage_explicit_last}
\end{align}
and the third is given by a unitary rotation on the last of these, $\sigma_{\pm1|2} = U_z(\pi/2)\sigma_{\pm|1}U^\dagger_z(\pi/2)$, where $U_z(\theta)$ denotes rotation about the $z$-axis through an angle $\theta$.
Note that
\begin{equation}
\rho_B = \frac{1}{2} \left( I +  \left( 1-\frac{2\alpha}{N} \right)\sigma_z \right).
\end{equation}

We begin by finding when each Bob's assemblage is steerable without consideration of Alice's reported non-detection events, in terms of the parameters $\alpha$ and $N$.
The purpose is to determine when the assemblage demonstrates steering, so appropriate ranges of these variables can be considered in the derivations to follow.
For this purpose, we can examine the steering inequality derived in main text Ref.~\cite{Jon11},
\begin{equation}
\frac{\Tr[\sigma_x \sigma_{+1|1}]}{p_{+1|1}} \leq \frac{1}{\sqrt{2}} \left( p_{+1|0} \sqrt{1-z_{+1|0}^2} + p_{-1|0} \sqrt{1-z_{-1|0}^2} \right).
\label{eq:app_spinning_top_inequality}
\end{equation}
The quantities appearing in this inequality can be calculated from the assemblage in Eqs.~\eqref{eq:app_assemblage_explicit_first}--\eqref{eq:app_assemblage_explicit_last} directly via $p_{a|x} \coloneqq \Tr[\sigma_{a|x}]$ and $z_{a|0} \coloneqq \Tr[\sigma_z \sigma_{a|0}]/p_{a|0}$.
Therefore, we find Eq.~\eqref{eq:app_spinning_top_inequality} is violated whenever 
\begin{equation}
\alpha \in \left( 0,\frac{2}{N+1} \right).
\label{eq:app_parameter_ranges}
\end{equation}
In this interval, $I$, we seek to derive an equation describing the behaviour of the cutoff-efficiency.

To do this, we will find an analytic solution to the SDP in Eq.~\eqref{eq:LTSDP} of the methods section.
We proceed in Section~\ref{app:sec_simplifications} by first simplifying this SDP into an equivalent form, by exploiting the symmetry present in the assemblages held by each Bob.
Using the same symmetries, its dual program is also given.
Then, we begin Section~\ref{sec:derive_primal_ansatz} by conducting extensive numerical tests to guide an ansatz for the primal variables which achieve $\epsilon^\star$, as a function of $\alpha$ and $N$ in $I$.
In doing so, we derive Eq.~\eqref{eq:3settings_analytic} of the main text.
We further discuss the equation in Section~\ref{sec:app_lower_bounds_on_cutoff}, providing tight lower bounds on it.
In Section~\ref{sec:derive_dual_ansatz}, we formulate an ansatz for the dual program.
This section of the SI concludes by showing that these ansatzes achieve the same value in Section~\ref{sec:check_optimality}, and therefore are optimal.

\subsection{Simplifications}
\label{app:sec_simplifications}

Recall that the set of probability distributions $\{ D(a |x,\lambda) \}_\lambda$ which encode the LHS constraints map $x$ to all outcomes, both null and non-null.
There are $3^3=27$ such distributions, and as many operators $\{\sigma_\lambda\}_\lambda$ over which the objective function must be maximized.
To reduce the dimension of this search space, we make some observations.

The first simplification arises from the rotational symmetry present in the assemblage.
Notice that the four assemblage members conditioned on settings 1 and 2 are all related by unitary rotations.
For instance, consider the Lie group $G$ generated by $\pi\hat{\sigma}_z/4$.
Each group element (represented by $k=0,\dots,3$) corresponds to a unitary operator $U_k \coloneqq  \exp[-i\pi k\hat{\sigma}_z/4]$. 
Let $\mathcal{U}_{k}(\bullet) \coloneqq U_k \bullet U_k^\dagger$.
Observe that the assemblage defined in Eqs.~\eqref{eq:app_assemblage_explicit_first}--\eqref{eq:app_assemblage_explicit_last} is $G$-covariant, meaning that $\{ \sigma_{a|x} \}_{a|x} \equiv \{\mathcal{U}_{k}(\sigma_{a|x})\}_{a|x} ~ \forall k.$
It is evident that the steered states conditioned on settings $x=1,2$ are related by
\begin{equation}
\sigma_{a'|x'} = \mathcal{U}_{k}(\sigma_{a|x}),
\label{eq:symmetry_in_steered_states}
\end{equation}
for an appropriate rotation determined by $k(a,x,a',x')$; the argument of this function will be omitted below for brevity.
%$k = [(x+x')+ 2(a+a')] \mod2$.
Moreover, the steered states conditioned on $x=0$ are invariant under these rotations.
This implies that the SDP constraint 
\begin{equation}
\sum_\lambda D(a |x,\lambda)\sigma_\lambda = \epsilon\sigma_{a|x},
\end{equation}
has interesting symmetry properties.
Suppose there exists, for all $x$, and all $a\neq\emptyset$, an ensemble $\{\sigma_\lambda\}_\lambda$  and a $\epsilon$ for which this constraint is satisfied. 
From Eq.~\eqref{eq:symmetry_in_steered_states}, one can then always construct a $G$-covariant ensemble $\{ \mathcal{U}_k \left(  \sigma_\lambda \right)/4 \}_{(\lambda,k)}$, which can exactly reproduce the $x=0$ steered states by the choice
\begin{equation}
\epsilon\sigma_{a|0} = \sum_{k,\lambda} D(a|0 ,\lambda)~ \mathcal{U}_k\left( \sigma_\lambda \right),
\label{eq:SI_invariant_LHS_constraints}
\end{equation}
and the $G$-covariant steered states by
% using Eq.~\eqref{eq:symmetry_in_steered_states} by
\begin{align}
\epsilon\sigma_{a'|x'} &=\mathcal{U}_{k}(\epsilon\sigma_{a|x}) \\
&= \mathcal{U}_{k}\left( \sum_\lambda D(a |x,\lambda)\sigma_\lambda \right) \\
% &= \sum_{\lambda} D(a' | x' ,\lambda)~ \sigma_\lambda \\
&= \sum_{k, \lambda} D(f(a,x,k) | g(x,k) ,\lambda)~ \mathcal{U}_k\left( \sigma_\lambda \right),
\label{eq:SI_covariant_LHS_constraints}
\end{align}
for $x'=1,2$ and $a'=\pm1$.
The functions $f$ and $g$ satisfy this equation for the choice
\begin{align}
a' = f(a,x,k) &\coloneqq 
\begin{cases}
(-1)^{h(x,k)} a &\mathrm{if} ~a = \pm 1, x\neq0\\
a &\mathrm{else},
\end{cases} \\
x' = g(x,k) &\coloneqq
\begin{cases}
(x\mod2) + 1, &k~\mathrm{odd}, x\neq0, \\
x &\mathrm{else}.
\end{cases}
\end{align}
Here we have defined the function $h(x,k) \coloneqq \lfloor ((k+x-1)\mod4)/2 \rfloor$, which has binary outputs.
Eqs.~\eqref{eq:SI_invariant_LHS_constraints} and \eqref{eq:SI_covariant_LHS_constraints} shows that all components of the assemblage (labelled by $a'$ and $x'$), can also be reproduced by the $G$-covariant ensemble $\{ \mathcal{U}_k \left(  \sigma_\lambda \right)/4\}_{(\lambda,k)}$.
This implies that, for the original problem, if we assume---without loss of generality---that the set $\{ \sigma_\lambda \}_\lambda$ is $G$-covariant, it is sufficient to confirm that, for any $\epsilon$, the elements $\sigma_{a|x}$ for $(a,x) \in \mathfrak{t}  \coloneqq \{(+1,0), (-1,0), (+1,1)\}$ are reproduced by it.
The set of tuples $\mathfrak{t}$ is minimal (but not unique) in this sense, in that it encodes a minimal set of constraints for the SDP, which are sufficient as a consequence of symmetry.

This symmetry can further be reflected in Alice's cheating strategies by partitioning them into equivalence classes (ECs).
We say that two different strategies with labels $\lambda, \lambda'$, are in the same equivalence class if 
\begin{equation}
D(a|x,\lambda) \equiv D(f(a,x,k)|g(x,k),\lambda')
\label{eq:SI_EC_definition}
\end{equation}
for some $k$.
Notice that these ECs have either $K$ or 1 members.
For brevity, we keep this multiplicity when referring to representatives of ECs with unit cardinality below, to simplify the writing of group averages over $G$.
For our problem, the original $27$ strategies can be partitioned into 9 ECs; these are characterized in Table \ref{table:appendix_ECs}.
This means that, instead of labelling strategies by $\lambda$, they can be represented by the tuple $(c,k)$, which uniquely specifies each strategy to be member $k$ of the EC $c$.
Moreover, instead of associating a different $\sigma_\lambda$ with each strategy, one can consider an operator $\sigma_c$ for each EC, with an appropriate rotation indexed by $k$, $\sigma_\lambda \rightarrow \mathcal{U}_k (\sigma_{c})$, corresponding to each strategy $(c,k)$. 
The optimization problem in Eq.~\eqref{eq:LTSDP} can therefore be expressed in a form which encodes these symmetries:
\begin{equation}
\begin{aligned}
\quad & \underset{\{\sigma_c\}_c}{\text{max}} & & \epsilon &\\
& \text{s. t.} & & \sum_{c,k} D(f(a,x,k) | g(x,k), c) \sigma_{c,k} = \epsilon\sigma_{a|x}, & a, x \in \mathfrak{t} \\ 
& & & \sum_{c,k} \sigma_{c,k} = \rho_B, & \\ 
& & & \sigma_{c} \geq 0,  & \forall c.
\end{aligned}
\label{eq:SI_symmetry_simplified_LTSDP}
\end{equation}

\begin{table}[htp]
\caption{Canonical representatives $\sigma_c$ for each of the nine equivalence classes of strategies, which comprise a LHS ensemble (if it exists).
These are found by enumerating the original $3^3=27$ strategies, and grouping them according into ECs Eq.~\eqref{eq:SI_EC_definition}.
The final three columns indicate the outcomes announced for each of the three settings.
The cardinality (number of strategies in each EC) are determined by having at least one non-null outcome announced for settings $x=1,2$.
}
\begin{center}
\begin{tabular}{|c||c|c|c|c|}
\hline
EC representative & Cardinality & $x=0$ & $x=1$ & $x=2$ \\
\hline\hline
$\sigma_0$ & 4 &+1 &+1 &+1 \\ 
$\sigma_1$ & 4 &+1 &+1 & $\emptyset$ \\ 
$\sigma_2$ & 1 & +1&$\emptyset$ &$\emptyset$ \\ 
$\sigma_3$ & 4 & -1& +1 & +1\\ 
$\sigma_4$ & 4 &-1 & +1&$\emptyset$ \\ 
$\sigma_5$ & 1 & -1& $\emptyset$& $\emptyset$\\ 
$\sigma_6$ & 4 &$\emptyset$ & +1 &+1 \\ 
$\sigma_7$ & 4 & $\emptyset$&+1 &$\emptyset$ \\ 
$\sigma_8$ & 1 &$\emptyset$ &$\emptyset$ &$\emptyset$ \\
\hline
\end{tabular}
\end{center}
\label{table:appendix_ECs}
\end{table}

The primal variables for the problem \eqref{eq:SI_symmetry_simplified_LTSDP} are the positive operators representative of each equivalence class, $\{\sigma_c\}_c$, and the real scalar $\epsilon$.
By introducing Hermitian operators $F_{a|x}, M,$ and $H_c$ as dual variables corresponding to the three types of constraints, we can form the Lagrangian
\begin{align}
\mathcal{L}%(\epsilon, \{F_{a|x}\}, N, \{H_\lambda\}) 
& = \epsilon + \Tr \left[ \sum\limits_{\substack{a,x \\ \in\mathfrak{t}}} F_{a|x}\left( \sum_{c,k} D(f(a,x,k) | g(x,k), c) \sigma_{c,k} - \epsilon\sigma_{a|x} \right)\right] \nonumber \\ 
&\qquad+ \Tr \left[M\left(\rho_B -  \sum_{c,k} \sigma_{c,k}\right)\right] + \Tr \left[\sum\limits_c H_c \sigma_c \right] \label{eq:app_lagrangian} \\
&= \Tr \left[ M \rho_B\right] + \epsilon \left(1 -  \Tr \sum\limits_{\substack{a,x \\ \in\mathfrak{t}}} F_{a|x} \right) \nonumber \\
&\qquad+ \Tr \left[ \sum\limits_c \sigma_c \left( \sum\limits_k \sum\limits_{\substack{a,x \\ \in\mathfrak{t}}} D(f(a,x,k) | g(x,k), c) U_k^\dagger F_{a|x} U_k + H_c - \sum\limits_k U_k^\dagger M U_k \right) \right]
\end{align}
The dual program is then
\begin{equation}
\begin{aligned}
& \text{min} & & \Tr M\rho_B &\\
& \text{s. t.} & & \sum\limits_{\substack{a,x \\ \in\mathfrak{t}}} \Tr F_{a|x}\sigma_{a|x} = 1 & \\
& & & \sum\limits_k \sum\limits_{\substack{a,x \\ \in\mathfrak{t}}} D(f(a,x,k) | g(x,k), c)U_k^\dagger F_{a|x} U_k  + H_c = \bar{M},& & \forall c, \\
\end{aligned}
\label{eq:app_symmetric_dual}
\end{equation}
where $\bar{M} \coloneqq  \sum_k U_k^\dagger M U_k$.
Notice that $H_c$ play the role of slack variables.
We explicitly include them in the dual formulation, because they will be important for deriving a certificate of optimality below.

\subsection{Constructing the ansatz: Primal}
\label{sec:derive_primal_ansatz}

The SDPs above can be straightforwardly solved, given fixed numeric values of $\alpha$ and $N$, using standard solvers.
We will go one step further, and derive a closed-form equation that explains the behaviour of $\epsilon^\star$ as a function of $\alpha$ and $N$, permitting us to take limits of the parameters analytically.
To do this, we will be guided by numerical simulations to construct primal and dual sets of variables, which are both valid (feasible) \emph{and} achieve the same values for the primal and dual objective functions, respectively.
That is, they are solutions with zero duality gap and are thus optimal.

\begin{figure}[htbp]
\includegraphics[width=\linewidth]{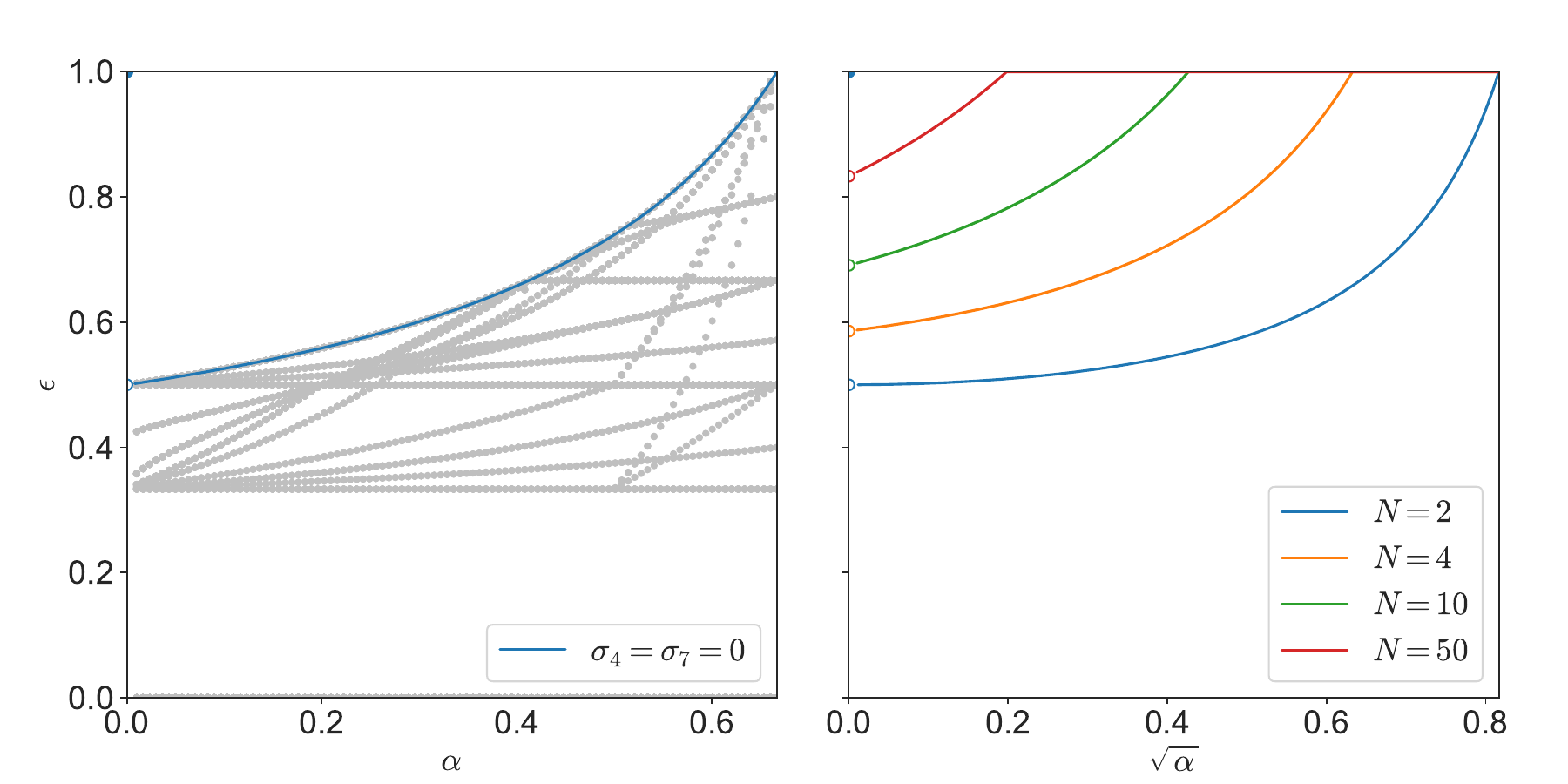}
\caption{ \textbf{Left:} Solutions $\epsilon^\star$ for $2^8$ families of SDP problems, defined by omitting all combinations of the matrices $\sigma_0,\dots,\sigma_7$ from the primal SDP formulation.
For each such problem, we fix $N=2$ and vary $\alpha$ to generate sets of grey points. 
The blue line corresponds to the solution for the sub-problem which encodes $\sigma_1=\sigma_4=\sigma_7=0$, which matches the solution giving the cutoff $\epsilon^\star$.
\textbf{Right:} Cutoffs from Eq.~\eqref{eq:3settings_analytic} for large numbers of Bobs, beyond the $N=2$ curve (reproduced from Fig~\ref{fig:main} here in blue).
This illustrates the remarkable observation that, for arbitrarily large $N$, there exists an interval of $\alpha$ such that $\epsilon^\star$ is below one, facilitating loss-tolerant steering.
Moreover, as shown in Section \ref{sec:app_lower_bounds_on_cutoff} of the SI, the smallest requirement on detector efficiency occurs in the limit $\alpha\rightarrow0$, where $\epsilon^\star = (1+\sqrt{2/N})^{-1}$, as reported in the main text.
}
\label{fig:plotting_extreme_points}
\end{figure}

The main difficulty is that the primal problem takes into account \emph{all} ECs of strategies for Alice, some of which are not used by her when the optimal value of $\epsilon^\star$ is obtained.
Here, and below, a $^\star$ denotes a primal or dual variable that is \emph{optimal}.
To construct an ansatz for primal variables, we first find the classes $c$ for which we can expect $\sigma_c^\star=0$.
To this end, we examine all families of solutions to \eqref{eq:SI_symmetry_simplified_LTSDP} for which subsets of the operators $\sigma_c$ are set to zero---or, equivalently, omitted from the problem formulation.
To begin, we observe that $\sigma_8$ represents the strategy for which null outcomes are alway announced, so we set it to zero.
For the remaining eight equivalence classes of strategies, there are $2^8 = 256$ such combinations to consider.
For each of these, we fix $N=2$ and perform a sweep over $0<\alpha\leq 2/3$, solving the SDP for each value of $\alpha$.
These results are shown by the grey points in the left sub-figure in Fig.~\ref{fig:plotting_extreme_points}.
The solution to the original problem \eqref{eq:SI_symmetry_simplified_LTSDP} is reproduced exactly by the simulations for which $\sigma_0=\sigma_4=\sigma_7=0$, and $\Tr[\sigma_i]>0$ for all others.

Many of the constraints in the primal problem \eqref{eq:app_symmetric_dual} are redundant, being satisfied from the symmetry properties encoded in the problem.
Defining group averages by $\bar{\mathcal{U}} \coloneqq \sum_k \mathcal{U}_k$, and removing the multiplicities on the ECs represented by $\sigma_2, \sigma_5,$ and $\sigma_8$, we can formulate the problem concisely:
\begin{equation}
\begin{aligned}
\text{max} \quad  & \epsilon \\
 \text{s. t.} \quad & \bar{\mathcal{U}}(\sigma_0) + \bar{\mathcal{U}}(\sigma_1) + \sigma_2 = \epsilon \sigma_{+|0} \\
& \bar{\mathcal{U}}(\sigma_3) + \bar{\mathcal{U}}(\sigma_4) + \sigma_5 = \epsilon \sigma_{-|0} \\
& \bar{\mathcal{U}}(\sigma_6) + \bar{\mathcal{U}}(\sigma_7) + \sigma_8 = (1-\epsilon) \rho_B \\
& \sum\limits_{i=0,3,6} \sigma_i + \mathcal{U}_1(\sigma_i) = \epsilon \sigma_{+|1} \\
 \forall c, \quad &  \sigma_{c} \geq 0 .
\end{aligned}
\label{eq:SI_minimal_primal_SDP}
\end{equation}
Let $\sigma_c \coloneqq  p_c(I + \hat{\bm{r}}_c \cdot\bm{\sigma})/2$ be a rank-one decomposition of the matrix variables in this problem, for some Bloch vector $\hat{\bm{r}}_c \coloneqq (\sin\theta_c \cos\phi_c, \sin\theta_c\sin\phi_c, \cos\theta_c)^T$.
With this parametrization, the final matrix inequality constraint is guaranteed provided $p_c \geq 0$.
It is evident from the left subfigure in Fig.~\ref{fig:plotting_extreme_points} that the solution to the original SDP for the case of $N=2$ Bobs occurs when we proceed by making the ansatz $\sigma_1=\sigma_4=\sigma_7=\sigma_8=0$.
Furthermore, we choose angles $\phi_3 = \phi_6=-\pi/4$, and $\theta_5 = \pi$.
The latter is a natural choice because $\sigma_5$ must be invariant under rotations generated by $\sigma_z$, and so must (when normalized) be located at the top or bottom of the Bloch sphere, meaning either $\theta_5=0$ or $\theta_5=\pi$.
We make the ansatz that this choice of primal variables will also be optimal for arbitrary $N$.
The constraints of the problem \eqref{eq:SI_minimal_primal_SDP} become
\begin{align}
4 p_0 + p_2 &= \epsilon (1-\alpha ) \label{eq:si4:con1}\\
4 p_3 + p_5 &=\epsilon \alpha \\
4 p_3 \cos \left(\theta _3\right) - p_5 &= \left(1-\frac{2}{N}\right) \epsilon \alpha \\ 
4 p_6 &=1-\epsilon \\ 
4 p_6 \cos \left(\theta _6\right) &= (1-\epsilon ) \left(1-\frac{2 \alpha }{N}\right) \\ 
\sqrt{2} p_3 \sin \left(\theta _3\right) + \sqrt{2} p_6 \sin \left(\theta _6\right) &= \frac{\sqrt{\alpha  (1-\alpha )} \epsilon }{N}\\ 
2 p_0 + 2 p_3 + 2 p_6 &= \frac{\epsilon }{2}\\ 
2 p_0 + 2 p_3 \cos \left(\theta _3\right)+2 p_6 \cos \left(\theta _6\right) &= \frac{\epsilon}{2} \left(1-\frac{2 \alpha }{N}\right) \label{eq:si4:con-1}
\end{align}
It is now straightforward to solve for the unknown variables by substitution.
We have
\begin{align}
p_6 &= \frac{1-\epsilon }{4} \\
\theta _6 &= \cos ^{-1}\left(1-\frac{2 \alpha }{N}\right),
\end{align}
and by rearranging, $p_5 = \epsilon \alpha -4 p_3$, $p_2 = \epsilon(1-\alpha)  - 4 p_0$ and $p_0 = \frac{1}{4} \left(2 \epsilon-4 p_3 -1\right)$.
Upon substituting, only three constraints remain:
\begin{align}
\frac{(N-1) \epsilon \alpha }{N} &= 2 p_3 \left(1 + \cos \left(\theta _3\right)\right), \label{eq:app_theta5_1} \\
2 N p_3 \sin \left(\theta _3\right) + (1-\epsilon) \sqrt{\alpha  (N-\alpha )} &= \epsilon \sqrt{2\alpha(1-\alpha)}, \\
\frac{\alpha  (2 \epsilon -1)}{N} &= 2 p_3 \left(1-\cos \left(\theta _3\right)\right) . \label{eq:app_theta5_2}
\end{align}
From \eqref{eq:app_theta5_1} we find
\begin{equation}
p_3 = \frac{(N-1) \epsilon \alpha }{2 N \left(1 + \cos \left(\theta _3\right)\right)},
\label{eq:app_p3_opt}
\end{equation}
which is valid if $\theta_3 \neq \pi$, and so Eq.~\eqref{eq:app_theta5_2} implies that
\begin{equation}
\theta_3 = \cos ^{-1}\left(\frac{N \epsilon -3 \epsilon +1}{N \epsilon +\epsilon -1}\right).
\label{eq:app_theta_3_opt}
\end{equation}
The last remaining constraint is 
\begin{align}
\alpha  (N-1) \epsilon  \tan \left(\frac{1}{2} \cos ^{-1}\left(\frac{(N-3) \epsilon +1}{N \epsilon +\epsilon -1}\right)\right) + (1-\epsilon) \sqrt{\alpha  (N-\alpha )} = \epsilon \sqrt{2\alpha(1-\alpha)}.
\end{align}
Using the identity $\tan(\frac{1}{2}\cos^{-1}(x/y))= \sqrt{\frac{y-x}{y+x}}$, this simplifies to 
\begin{equation}
\epsilon \sqrt{2\alpha(1-\alpha)} + (\epsilon -1) \sqrt{\alpha  (N-\alpha )} = \alpha  \sqrt{2 \epsilon -1} \sqrt{(N-1) \epsilon },
\label{eq:primal_implicit_equation}
\end{equation}
which implicitly defines a solution for the remaining variable $\epsilon$.
This can be converted into a quadratic equation in $\epsilon$, from which we can find a closed-form solution.
This solution exactly matches the numerics and is given by $\epsilon^{\star}(\alpha,N) = e_{\alpha,N}$, where
\begin{equation}
e_{\alpha,N} \coloneqq \frac{2N -\alpha  (N+1)+2 \sqrt{2(1-\alpha)} \sqrt{N-\alpha }+\sqrt{\alpha 
   (N-1) \left( (\alpha +4) N -5 \alpha -4 \sqrt{2(1- \alpha) }
   \sqrt{N-\alpha }\right)}}{2 \left(2
   \sqrt{2(1-\alpha)} \sqrt{N-\alpha }+N+2-\alpha -2 \alpha  N\right)},
\label{eq:3settings_analytic_sup}
\end{equation}
as given in the main text.

\subsection{Tight lower bounds on $e_{\alpha,N}$}
\label{sec:app_lower_bounds_on_cutoff}

Here, we provide tight lower bounds on $e_{\alpha,N}$, for two reasons.
The first is to justify the statement regarding the minimum efficiency being obtained in the singular limit $\alpha \rightarrow 0$.
Secondly, will require $e_{\alpha,N}>1/2$ for the ansatz of the dual variables to be well-defined below.

Recall that we are interested in the interval $I \coloneqq \alpha \in (0, 2/(N+1))$.
To obtain a lower bound on $e_{\alpha,N}$ in $I$, which is tight for any integer $N\geq 2$, we first observe that it is monotonically increasing.
To this end, define
\begin{align}
z_0 &\coloneqq 2N -\alpha  (N+1)+2 \sqrt{2(1-\alpha)} \sqrt{N-\alpha }, \\
z_1 &\coloneqq \sqrt{\alpha (N-1) \left( (\alpha +4) N -5 \alpha -4 \sqrt{2(1- \alpha) }\sqrt{N-\alpha }\right)}, \\
z_2 &\coloneqq 2 \left(2 \sqrt{2(1-\alpha)} \sqrt{N-\alpha }+N+2-\alpha -2 \alpha  N\right),
\end{align}
so that $e_{\alpha,N} = (z_0 + z_1)/z_2$.
We will show that both $z_0/z_2$ and $z_1/z_2$ are strictly increasing functions of alpha in $I$, for any integer $N\geq2$.

First, we observe that
\begin{align}
\frac{\partial}{\partial\alpha}\frac{z_0}{z_2} = \frac{(N-1) \left(-\sqrt{2} \alpha  (N+4)+3 \sqrt{1-\alpha } N \sqrt{N-\alpha }+2 \sqrt{1-\alpha } \sqrt{N-\alpha }+\sqrt{2} (3 N+2)\right)}{2 \sqrt{1-\alpha } \sqrt{N-\alpha } \left(-\alpha -2 \alpha  N+2 \sqrt{2-2 \alpha } \sqrt{N-\alpha }+N+2\right)^2}.
\end{align}
The denominator is positive in $I$.
To see that the numerator is also positive, by omitting some strictly positive terms, we know that it is lower bounded by $\sqrt{2} (3 N+2)-\sqrt{2} \alpha  (N+4)>0$.
Hence, $z_0/z_2$ is strictly increasing in $I$.

Now, for the second term, note that
\begin{equation}
\frac{\partial z_2}{\partial\alpha} = 2 \left(-\frac{\sqrt{2-2 \alpha }}{\sqrt{N-\alpha }}-\frac{2 \sqrt{N-\alpha }}{\sqrt{2-2 \alpha }}-2 N-1\right),
\end{equation}
which implies that $z_2^{-1}$ is strictly increasing (and positive).
For the final term, note that
\begin{equation}
\frac{\partial z_1^2}{\partial\alpha} = 2 (N-1) \left[(\alpha +2) N+\frac{\sqrt{2} \alpha  (3-4 \alpha )+\sqrt{2} (3 \alpha -2) N-5 \sqrt{1-\alpha } \alpha  \sqrt{N-\alpha }}{\sqrt{1-\alpha } \sqrt{N-\alpha }}\right].
\end{equation}
The term in square brackets is strictly positive for $N\geq2$ and $0<\alpha<1$, and which contains the interval of  interest. 
This means that its square root, $z_1$, is strictly positive and increasing too.
We conclude that $z_1/z_2$ must also be strictly increasing in $I$.

Now, the sum of any two strictly increasing functions is also strictly increasing.
Therefore $e_{\alpha,N} = (z_0 + z_1)/z_2$ is strictly increasing in $I$.
This implies that, for all values of $N$, $\min_{\alpha \in I} e_{\alpha,N}$ is achieved at the start of the interval.
That is,
\begin{align}
e_{\alpha,N} &> \lim\limits_{\alpha\rightarrow0^+} e_{\alpha,N} \\
&= \frac{1}{1 + \sqrt{\frac{2}{N}}}.
\end{align}
This also shows that $e_{\alpha,N}>1/2$ for all valid $N$.
Finally, we observe that this limit is singular, since for $\alpha=0$ the family entangled states in \eqref{eq:mainclass} of the main text are unentangled, therefore non-steerable, and so $\epsilon^\star=1$.

\subsection{Constructing the ansatz: Dual}
\label{sec:derive_dual_ansatz}

Note that have not yet shown that this choice of primal variables (and hence the solution) is \emph{optimal}.
The set of primal variables achieving the objective value in Eq.~\eqref{eq:3settings_analytic} above are only \emph{one} feasible set of variables, i.e. they satisfy the constraints of the primal problem.
However, we can prove optimality by analyzing the dual program and constructing a set of dual variables for which the dual objective function is equal to the primal objective function above.
In other words, they will form a primal/dual pair with zero duality gap and are thus optimal.
That is, the set of dual variables we construct will provide a certificate of optimality for this ansatz.

For convenience, we summarize the primal variables forming the ansatz above:
\begin{align}
\sigma_0 &= \frac{\alpha  (1- e_{\alpha,N})+(2-\alpha ) N e_{\alpha,N}  -N}{4 N} \ketbra{0} \\
\sigma_1 &= 0 \\
\sigma_2 &=  \frac{(1-e_{\alpha,N}) (N-\alpha )}{N} \ketbra{0} \\
\sigma_3 &=  \frac{\alpha  (N-1) e_{\alpha,N} }{2 N \left(1+ \cos \theta_3\right)} \cdot \frac{1}{2}\left( I + \hat{\bm{r}}_3 \cdot \bm{\sigma} \right) \label{eq:SM_sig3_anz}\\
\sigma_4 &= 0 \\
\sigma_5 &=  \alpha  e_{\alpha,N}  \left(1+\frac{2(1-N)}{N(1+\cos\theta_3)}\right) \ketbra{1}\\
\sigma_6 &=  \frac{1-e_{\alpha,N} }{4} \cdot \frac{1}{2}\left( I + \hat{\bm{r}}_6 \cdot \bm{\sigma} \right) \label{eq:SM_sig6_anz}\\
\sigma_7 &= 0 \\
\sigma_8 &= 0
\end{align}
where the Bloch vectors $\hat{\bm{r}}_i \coloneqq (\sin\theta_i \cos\phi_i, \sin\theta_i\sin\phi_i, \cos\theta_i)^T$ are defined by
\begin{align}
\theta_3 &= \cos^{-1}\left( \frac{(N-3) e_{\alpha,N} +1}{(N+1) e_{\alpha,N} -1} \right), \label{eq:primal_ansatz_theta3}\\
\phi_3 &= -\frac{\pi}{4}, \\
\theta_6 &= \cos^{-1}\left( 1- \frac{2\alpha}{N} \right), \\
\phi_6 &= -\frac{\pi}{4}.
\end{align}

We first remove some unconstrained degrees of freedom in the dual program in \eqref{eq:app_symmetric_dual}, by using the fact that the steered state $\sigma_{+|z}/\Tr[\sigma_{+|z}]$ is pure.
Since pure states are extreme points in the space of two-qubit density matrices, this implies that the portion of the LHS ensemble which averages to it must also consist of the same pure state, $\sigma_i = p_i \ketbra{0}$ for $i=0,1,2$.
Using this fact, we can formulate \eqref{eq:app_symmetric_dual} into the equivalent dual program:
\begin{align}
\text{min}\quad & \Tr M\rho_B \\
\text{s. t.}\quad & \Tr \left[ F_{+|0}\sigma_{+|0} + F_{-|1}\sigma_{-|1} + 4 F_{+|1}\sigma_{+|1} \right] = 1 \label{app_dual_constraint_scalar}\\
 & \ev{M}{0} = \bra{0} \left(F_{+|0} + F_{+|1} + \mathcal{U}^\dagger_1(F_{+|1}) \right) \ket{0} + h_0 & \label{app_dual_constraint_H0} \\
& \ev{M}{0} = \bra{0} \left(F_{+|0} + F_{+|1}\right) \ket{0} + h_1 & \label{app_dual_constraint_H1} \\
 & \ev{M}{0} = \bra{0} F_{+|0} \ket{0} + h_2 & \label{app_dual_constraint_H2} \\
 & M = F_{-|1} + F_{+|1} + \mathcal{U}^\dagger_1(F_{+|1}) + H_3 & \label{app_dual_constraint_H3}\\
 & M = F_{-|1} + F_{+|1} +  H_4 & \label{app_dual_constraint_H4} \\
 & M = F_{-|1} +  H_5 & \label{app_dual_constraint_H5} \\
 & M = F_{+|1} + \mathcal{U}^\dagger_1(F_{+|1}) + H_6 &  \label{app_dual_constraint_H6}\\
 & M = F_{+|1} + H_7 & \label{app_dual_constraint_H7} \\
 & M =  H_8 & \label{app_dual_constraint_H8} \\
\mathrm{for} ~c=0,1,2, \quad  &  h_c \geq 0 \\
\mathrm{for}~c=3,\dots,8, \quad  &  H_c \geq 0.
\end{align}

The variables in the dual program are the Hermitian matrices $M, F_{+|0}, F_{-|1}, F_{+|1}$, three real scalars $h_i$ and the six positive-semidefinite matrices $H_c$.
Motivated by numerical simulations, and examining the structure of the dual constraints, we formulate an ansatz now for the dual variables.
We begin by assuming the first four of these matrices are of the form
\begin{align}
M &= 
\left(
\begin{array}{ccc}
\mu_0 & 0 \\
0  & \mu_1
\end{array}
\right), \\
F_{+|0} &= \mu_0 \ketbra{0}, \\
F_{-|1} &= 
\left(
\begin{array}{ccc}
c_0 & 0 \\
0  & \mu_1
\end{array}
\right), \\
F_{+|1} &= c_1 I + c_2\sigma_x - c_1\sigma_z
\end{align}
for unknowns $\mu_0, \mu_1, c_0, c_1, c_2$. 
The structure of the $H_c$'s will be guided by the so-called \emph{complementary slackness} condition for optimality.
These are necessary conditions that a primal/dual set of optimal variables must satisfy.
Following Boyd and Vandenberghe \cite{Boyd04} (\S~5.5.2), for any primal and dual optimal set of variables that have zero duality gap, we know that
\begin{align}
\epsilon^\star &= \Tr [M^\star \rho_B] \\
&\geq \epsilon^\star + \Tr \left[ \sum\limits_{\substack{a,x \\ \in\mathfrak{t}}} F^\star_{a|x}\left( \epsilon^\star\sigma_{a|x}  - \sum_{c,k} D(f(a,x,k) | g(x,k), c) \sigma^\star_{c,k} \right)\right] + \Tr \left[M^\star\left(\rho_B - \sum_{c,k} \sigma^\star_{c,k} \right)\right] + \Tr \left[\sum\limits_\lambda H^\star_\lambda \sigma^\star_\lambda\right] \label{app:eq_starred_lagrangian} \\
&\geq \epsilon^\star.
\end{align}
The second line follows from the definition of $\mathcal{L}$ in \eqref{eq:app_lagrangian}, and the third from the facts that primal feasibility holds for the primal variables, and $\Tr[XY] \geq 0$ for any two positive semidefinite matrices $X,Y$.
It can therefore be deduced that both inequalities in this chain hold with equality.
This implies that the final term in \eqref{app:eq_starred_lagrangian} vanishes,
\begin{equation}
\sum\limits_\lambda \Tr\left[ H^\star_\lambda \sigma^\star_\lambda\right] = 0.
\end{equation}
Since each term in this sum is non-negative, all terms must vanish, and so for any primal/dual optimal set of variables, the ranges of $H^\star_\lambda$ and $\sigma^\star_\lambda$ must be pair-wise orthogonal.
This implies we should express the dual variables $H^\star_c$ in the form
\begin{equation}
H^\star_c \coloneqq h_c \left(I - \frac{\sigma_c^\star}{\Tr[\sigma_c^\star]} \right)
\end{equation}
when $\sigma_c^\star$ is not zero, and where $h_\lambda$ are (undetermined) positive real scalars.

In particular, from Eqs.\eqref{eq:SM_sig3_anz} and \eqref{eq:SM_sig3_anz} we define
\begin{align}
H_3 &= \frac{h_3}{2}\left( I - \frac{\sin{\theta_3}}{\sqrt{2}}(\sigma_x - \sigma_y)  + \cos{\theta_3} \sigma_z \right), \\
H_6 &= \frac{h_6}{2}\left( I - \frac{\sin{\theta_6}}{\sqrt{2}}(\sigma_x - \sigma_y)  + \cos{\theta_6} \sigma_z \right).
\end{align}
The other remaining $H_c$'s are not important for deriving the dual optimal variables, only their existence as positive semidefinite matrices is important from the point of view of dual feasibility; we confirm this at the end.

The dual constraints which allow the unknowns to be determined are Eqs.~\eqref{app_dual_constraint_H3} and \eqref{app_dual_constraint_H6}:
\begin{align}
F_{-|1} + F_{+|1} + \mathcal{U}^\dagger_1(F_{+|1}) + H_3 &= M, \label{eq:app_minus1multiplier}\\
F_{+|1} + \mathcal{U}^\dagger_1(F_{+|1}) + H_6 &= M. \label{eq:app_plus1multiplier}
\end{align}
The latter of these, Eq.~\eqref{eq:app_plus1multiplier}, expressed in terms of the Pauli operators reads
\begin{align}
\left ( 2 \left(2 c_1+h_6\right) \right) I + &
\left (2 c_2-\frac{2 \sqrt{2} h_6 \sqrt{\alpha  (N-\alpha )}}{N}  \right) (\sigma_x -\sigma_y) -
\left (4 c_1+\frac{2 h_6 (N-2 \alpha )}{N}  \right) \sigma_z \\
&= (\mu _0+\mu _1)I + (\mu _0-\mu _1) \sigma_z
\end{align}
Matching off-diagonal (in the $\sigma_z$-basis) terms, we find
\begin{equation}
h_6 =  \frac{c_2 N}{\sqrt{2\alpha  (N-\alpha )}},
\end{equation}
from which matching the identity and $\sigma_z$ terms implies
\begin{align}
c_1 &= \frac{1}{4} \left(\mu _0+\mu _1- \frac{\sqrt{2} c_2 N}{\sqrt{\alpha  (N-\alpha )}}\right), \\
c_2 &= \frac{\mu _0 \sqrt{\alpha  (N-\alpha)}}{\sqrt{2} \alpha }. \label{app_c2_val}
\end{align}
Similarly, Eq.~\eqref{eq:app_minus1multiplier} is
\begin{align}
\left( c_0+4 c_1+2 h_3+\mu _1 \right) I + &
\left( 2 \left(c_2-\frac{h_3 \sin\theta_3}{\sqrt{2}} \right) \right) (\sigma_x -\sigma_y) -
\left( c_0-4 c_1-2 h_3 z_3-\mu _1 \right) \sigma_z \\
&= (\mu _0+\mu _1)I + (\mu _0-\mu _1) \sigma_z.
\end{align}
Once again matching terms, we find
\begin{align}
h_3 &= \frac{\sqrt{2}c_2}{\sin\theta_3} \label{app_h3_val} \\
c_0 &= \frac{\mu _0 \left(N-2 \csc \left(\theta_3\right) \sqrt{\alpha  (N-\alpha )}\right)}{\alpha }-\mu _1 \\
\mu_1 &= \frac{\mu _0 \left( N-\alpha +  \left(1 + \cos \left(\theta_3\right)\right) \csc \left(\theta_3\right) \sqrt{\alpha  (N-\alpha )}\right)}{\alpha}
\end{align}
Using the identity $\left( 1 + \cos \left(\theta_3\right)\right) \csc \left(\theta_3\right) = \sqrt{2(1+\cos \left(\theta_3\right))/(1- \cos \left(\theta_3\right))}$, and the definition of $\theta_3$ in Eq.~\eqref{eq:primal_ansatz_theta3}, we can express the equation for $\mu_1$ in the form
\begin{equation}
\mu _1= \frac{\mu _0 \left( 2 (N-\alpha) - \frac{2 \sqrt{e_{\alpha,N} (N-1)} \sqrt{\alpha  (N-\alpha )}}{\sqrt{2 e_{\alpha,N}-1}}\right)}{2 \alpha }.
\end{equation}
This only remaining degree of freedom is $\mu_0$, which we choose to satisfy the scalar constraint in Eq.~\eqref{app_dual_constraint_scalar},
\begin{equation}
\Tr \left[ F_{+|0}\sigma_{+|0} + F_{-|1}\sigma_{-|1} + 4 F_{+|1}\sigma_{+|1} \right] = 1.
\end{equation}
For the choice of operators we have made, this reads
\begin{equation}
(1-\alpha ) \mu _0+4 \left(-\frac{1}{2} c_1 \left(1-\frac{2 \alpha }{N}\right)+\frac{\sqrt{(1-\alpha ) \alpha } c_2}{N}+\frac{c_1}{2}\right)+\alpha  c_0 \left(1-\frac{1}{N}\right)+\frac{\alpha  \mu _1}{N} = 1,
\end{equation}
from which we deduce that
\begin{equation}
\mu _0 = \frac{\alpha  e_{\alpha,N} N}{2 e_{\alpha,N} \left(\alpha(N-\alpha) -\frac{2 \alpha  \sqrt{e_{\alpha,N} (N-1)} \sqrt{\alpha  (N-\alpha )}}{\sqrt{2 e_{\alpha,N}-1}} + \sqrt{2} \sqrt{(1-\alpha) \alpha } \sqrt{\alpha  (N-\alpha )}\right)+\frac{\alpha  \sqrt{e_{\alpha,N} (N-1)} \sqrt{\alpha  (N-\alpha )}}{\sqrt{2 e_{\alpha,N}-1}}}.
\end{equation}
This defines all relevant degrees of freedom in our construction for the dual variables.
It remains to check (for dual feasibility) that the slack variables are valid for this choice; $h_c\geq0$ for $i=0,1,2$, and $H_c\geq 0$ for $i=3,\dots,8$.
These matrices are defined through Eqs.~\eqref{app_dual_constraint_H0}--\eqref{app_dual_constraint_H8}.
For the scalars, we have
\begin{align}
h_0 &= \ev{\left( M - F_{+|0} - F_{+|1} - \mathcal{U}^\dagger_1(F_{+|1})\right)}{0} \\
&=0 \\
h_1 &= \ev{\left( M - F_{+|0} - F_{+|1}\right)}{0} \\
&=0 \\
h_2 &= \ev{\left( M - F_{+|0}\right)}{0} \\
&=0.
\end{align}
We now check the eigenvalues of the $H_c$ matrices are all non-negative.
%H_3
The eigenvalues of
\begin{equation}
H_3 = \frac{h_3}{2}\left( I - \frac{\sin{\theta_3}}{\sqrt{2}}(\sigma_x - \sigma_y)  + \cos{\theta_3} \sigma_z \right)
\end{equation}
are $0$ and $h_3$. 
From \eqref{app_h3_val} and \eqref{app_c2_val} we know that 
\begin{equation}
h_3 = \frac{\mu _0 \sqrt{\alpha  (N-\alpha)}}{\alpha \sin\theta_3} > 0,
\end{equation}
where the inequality follows from observing that $\mu_0>0$, multiplies a term that is strictly positive in the interval $I$.
%H_4
The least eigenvalue of $H_4 = M - F_{-|1} - F_{+|1}$ is
\begin{equation}
\frac{1}{2} \left(\mu _0 -c_0-2 c_1-\sqrt{\left(c_0+2 c_1-\mu _0\right)^2-4 \left(2 c_0 c_1 -2 c_1 \mu _0-c_2^2\right)}\right),
\end{equation}
This evaluates to zero; to see this, observe that 
\begin{align}
&2 c_0 c_1 - 2 c_1 \mu _0-c_2^2 = \\ 
&\frac{\mu _0^2 (N-\alpha ) \left[\alpha  (1-2e_{\alpha,N}) \sqrt{e_{\alpha,N} (N-1)}+\sqrt{2 e_{\alpha,N}-1} \left(\sqrt{\alpha  (N-\alpha )}-e_{\alpha,N} \left(\sqrt{2(1-\alpha) \alpha )}+\sqrt{\alpha  (N-\alpha )}\right)\right)\right]}{2 \alpha ^2 (2e_{\alpha,N}-1) \sqrt{e_{\alpha,N}(N-1)}},
\end{align}
which evaluates to zero.
This is because $e_{\alpha,N}$ is a solution for $\epsilon$ in Eq.~\eqref{eq:primal_implicit_equation}, which upon rearranging and substituting causes the term in the square brackets to vanish.
%H_5
Now, $H_5 = M - F_{-|1} = \mathrm{diag}(\mu_0 - c_0, 0)$, which has non-negative eigenvalues since
\begin{equation}
\mu_0 - c_0 = \frac{\sqrt{2e_{\alpha,N}-1} \mu _0 \sqrt{\alpha  (N-\alpha )}}{\alpha  \sqrt{e_{\alpha,N}(N-1)}},
\end{equation}
which is strictly positive in $I$.
%H_6
Similarly, the eigenvalues of 
\begin{equation}
H_6 = \frac{h_6}{2}\left( I - \frac{\sin{\theta_6}}{\sqrt{2}}(\sigma_x - \sigma_y)  + \cos{\theta_6} \sigma_z \right)
\end{equation}
are $0$ and $h_6$, the latter of which is positive because $h_6 = \mu_0 N/(2\alpha) > 0$.
%H_7
Both eigenvalues of $H_7$ are positive.
To see this, the least eigenvalue of $H_7 = M - F_{+|1}$ is 
\begin{equation}
\frac{1}{2} \left(-\sqrt{\left(2 c_1-\mu _0-\mu _1\right)^2-4 \left(-2 c_1 \mu _0-c_2^2+\mu _0 \mu _1\right)}-2 c_1+\mu _0+\mu _1\right).
\end{equation}
To check that this eigenvalue is positive, it suffices to verify that
\begin{equation}
\left(-2 c_1+\mu _0+\mu _1\right){}^2>\left(2 c_1-\mu _0-\mu _1\right){}^2-4 \left(-2 c_1 \mu _0-c_2^2+\mu _0 \mu _1\right),
\end{equation}
which is equivalent to 
\begin{equation}
\mu _0 \left(\mu _1-2 c_1\right)>c_2^2.
\end{equation}
Upon substitution, this becomes 
\begin{equation}
\alpha ^2 e_{\alpha,N}(2e_{\alpha,N}-1) \mu _0^2 \left(\alpha ^2-\alpha  N+\sqrt{2} \sqrt{-((\alpha -1) \alpha )} \sqrt{\alpha  (N-\alpha )}\right)<0,
\end{equation}
which is valid in $I$ because $e>1/2$, $\mu_0>0$.
%H_8
Finally, since the eigenvalues of $M=H_8$ are both strictly positive, we conclude that $H_8$ is also strictly positive.
Hence, all dual variables for the ansatz we have made satisfy the constraints of the dual program. 

\subsection{Certifying optimality}
\label{sec:check_optimality}

It remains to see that this choice implies a zero duality gap for the values achieved by the primal and dual objective functions.
The duality gap for our ansatz is given by
\begin{equation}
\Tr\left[M \rho_B \right] - e_{\alpha,N} = \mu _0 \left(1-\frac{\alpha }{N}\right)+\frac{\alpha  \mu _1}{N} - e_{\alpha,N}.
\label{eq:duality_gap_for_ansatz}
\end{equation}
Since we know that $e_{\alpha,N}$ defines a solution for $\epsilon$ in Eq.~\eqref{eq:primal_implicit_equation}, we know, by rearranging, that it also satisfies
\begin{equation}
\frac{\alpha  \sqrt{e_{\alpha,N} (N-1)}}{\sqrt{2 e_{\alpha,N}-1}} = \frac{\sqrt{2} \sqrt{(1-\alpha) \alpha } e_{\alpha,N}+(e_{\alpha,N}-1) \sqrt{\alpha  (N-\alpha )}}{2e_{\alpha,N}-1}.
\end{equation}
We can then simplify the eigenvalues of $M$ into the forms
\begin{equation}
\mu_1 = -\frac{e_{\alpha,N} \mu _0 \left(\alpha- N+\sqrt{2} \sqrt{1-\alpha } \sqrt{(N-\alpha )}\right)}{\alpha (2 e_{\alpha,N}-1)}
\end{equation}
and
\begin{equation}
\mu_0 = \frac{\alpha e_{\alpha,N} (2 e_{\alpha,N}-1) N}{\alpha ^2 (1-3 e_{\alpha,N})+\alpha  (3e_{\alpha,N}-1) N-\sqrt{2} e_{\alpha,N} \sqrt{(1-\alpha) \alpha } \sqrt{\alpha  (N-\alpha )}}.
\end{equation}
Upon substituting into Eq.\eqref{eq:duality_gap_for_ansatz}, it is straightforward to verify that
\begin{equation}
\Tr\left[M\rho_B \right] - e_{\alpha,N} = 0
\end{equation}
for the sets of primal and dual variables, we have defined above, implying that they form an optimal primal and dual pair.
That is, we know that Eq.~\eqref{eq:3settings_analytic} from the main text is a closed-form solution to the optimization problem and hence represents the cutoff efficiencies for each Bob's assemblage.

\section{Hypothesis testing}\label{app:hypotest}

Denote $\vec{\theta}$ the vector of all the unknown parameters to be estimated, and $D$ the experiment data obtained. Suppose there are altogether $R$ different measurement settings, accounting for different POVM groups and different initial states. The $r$-th POVM group has a total measurement shots $S_r$ for $O_r$ different outcomes, with the occurrence statistics $\vec{D}_r:=(D_{r1},D_{r2},...,D_{rO_r})$. Hence, $S_r=\sum_{i=1}^{O_r}D_{ri}$.

Given $D$, one can use certain algorithm to obtain an estimation $\hat{\vec{\theta}}$. Then the natural question is, to what degree is $\hat{\vec{\theta}}$ consistent with $D$? This is usually answered by performing {\textit{hypothesis testing}} in statistics. More specifically, in our measurement-based scenario, we need to do a {\textit{multinomial test}}~\cite{multinomial} as follows.

We would like to test the null hypothesis $H_0:\vec{\theta}=\hat{\vec{\theta}}$. For the $r$-th measurement group, this hypothesis will give a prediction of the normalised measurement probabilities $\vec{f}_r=(f_{r1},f_{r2},...,f_{rO_r})$ where $\sum_{i=1}^{O_r}f_{ri}=1$. Then the probability (likelihood) of observing $\vec{D}_r$ under the null hypothesis is clearly $$p(\vec{D}_r|H_0)=S_r!\prod_{i=1}^{O_r}\frac{f_{ri}^{D_{ri}}}{D_{ri}!}.$$ One then typically define an {\textit{alternative hypothesis}} $H_1$ where the predicted measurement probabilities, instead of being $\vec{f}_r$, should be $(p_{r1}^{{\rm{MLE}}},p_{r2}^{\rm{MLE}},...,p_{rO_r}^{\rm{MLE}})$ where $p_{ri}^{\rm{MLE}}=D_{ri}/S_r$ is the maximum likelihood estimate from data. The probability of observing $\vec{D}_r$ under the alternative hypothesis is then $$p(\vec{D}_r|H_1)=S_r!\prod_{i=1}^{O_r}\frac{(p_{ri}^{\rm{MLE}})^{D_{ri}}}{D_{ri}!}.$$ Now a likelihood ratio test can be done as ${\rm{LRT}}_r:=p(\vec{D}_r|H_0)/p(\vec{D}_r|H_1)$. Assume $H_0$ is true, then asymptotically ($S_r\rightarrow\infty$) the distribution of $-2\ln({\rm{LRT}}_r)=-2\sum_{i=1}^{O_r}D_{ri}\ln(f_{ri}/p_{ri}^{\rm{MLE}})$ converges to the $\chi^2$ distribution with $O_{r}-1$ degrees of freedom. Since measurements from different measurement settings are independent, we can employ the additivity of $\chi^2$ distribution to assert that $\sum_{r=1}^R -2\ln ({\rm{LRT}}_r) =-2\sum_{r=1}^R\sum_{i=1}^{O_r}D_{ri}\ln(f_{ri}/p_{ri}^{\rm{MLE}})$ should asymptotically follow the $\chi^2$ distribution with $\sum_{r=1}^R (O_r-1)$ degrees of freedom, when $H_0$ holds. One can thus select a significance level $s_\alpha$ to decide whether $H_0$ will be rejected, based on the calculated value of $\sum_{r=1}^R -2\ln({\rm{LRT}}_r)$.

For our assemblage tomography part, $\sum_{r=1}^R (O_r-1)=\sum_{r=1}^{{108}}(6-1)=540$ and its significance level $s_\alpha=5\%$ corresponds to the critical region $\sum_{r=1}^R -2\ln ({\rm{LRT}}_r)\geq 595.1683$, while our tomography result gives $\sum_{r=1}^R -2\ln ({\rm{LRT}}_r)=572.8553$. Hence the null hypothesis is accepted, and our tomography result matches the data.

\end{document}